\documentclass{aa} 
\usepackage{natbib}
\usepackage{graphicx}
 \usepackage[varg]{txfonts}
\usepackage{natbib}
\usepackage[colorlinks=true,citecolor=blue,linkcolor=blue]{hyperref}
\usepackage{gensymb}
\usepackage{subcaption}
\usepackage{lscape}

\graphicspath{{./figs/}}

\newcommand{\nustar}{\textsl{NuSTAR}\xspace}

\usepackage{footmisc}
\usepackage[dvipsnames]{xcolor}

\begin{document}

\title{Continuum, cyclotron line, and absorption variability in the high-mass X-ray binary Vela X-1}

\author{
C.~M.~Diez\inst{1}\and
V.~Grinberg\inst{2,1}\and
F.~F\"urst\inst{3} \and
E.~Sokolova-Lapa\inst{4,5} \and
A.~Santangelo\inst{1}\and
J.~Wilms\inst{4}\and
K.~Pottschmidt\inst{6,7}\and
S.~Mart\'inez-N\'u\~nez\inst{8}\and
C.~Malacaria\inst{9}\and
P.~Kretschmar\inst{10}
}

\institute{Institut f\"ur Astronomie und Astrophysik, Universität T\"ubingen, Sand 1, 72076 T\"ubingen, Germany \\
\texttt{diez@astro.uni-tuebingen.de}
\and
European Space Agency (ESA), European Space Research and Technology Centre (ESTEC), Keplerlaan 1, 2201 AZ Noordwijk, The Netherlands
\and
Quasar Science Resources S.L for European Space Agency (ESA), European Space Astronomy Centre (ESAC), Camino Bajo del Castillo s/n, 28692 Villanueva de la Cañada, Madrid, Spain
\and 
Dr.~Karl Remeis Sternwarte \& Erlangen Centre for Astparticle Physics, Friedrich-Alexander-Universit\"at Erlangen-N\"urnberg, Sternwartstr.~7, 96049 Bamberg, Germany
\and
Sternberg Astronomical Institute, M.~V.~Lomonosov Moscow State University,
Universitetskij pr., 13, Moscow 119992, Russia
\and
CRESST, Department of Physics, and Center for Space Science and Technology, UMBC, Baltimore, MD 21250, USA
\and
NASA Goddard Space Flight Center, Astrophysics Science Division, Greenbelt, MD 20771, USA
\and 
Instituto de F\'isica de Cantabria (CSIC-Universidad de Cantabria), E-39005, Santander, Spain
\and
Universities Space Research Association, Science and Technology Institute, 320 Sparkman Drive, Huntsville, AL 35805, USA
\and
European Space Agency (ESA), European Space Astronomy Centre (ESAC), Camino Bajo del Castillo s/n, 28692 Villanueva de la Cañada, Madrid, Spain  
}

\date{Received 9 July 2021 / Accepted 23 December 2021}

\abstract{
Because of its complex clumpy wind, prominent cyclotron resonant scattering features, intrinsic variability, and convenient physical parameters (close distance, high inclination, and small orbital separation), which facilitate the observation and analysis of the system, Vela X-1 is one of the key systems for understanding accretion processes in high-mass X-ray binaries on all scales.
We revisit Vela X-1 with two new observations taken with \nustar at orbital phases $\sim$0.68--0.78 and $\sim$0.36--0.52, which show a plethora of variability and allow us to study the accretion geometry and stellar wind properties of the system.
We follow the evolution of spectral parameters down to the pulse period timescale using a partially covered power law continuum with a Fermi-Dirac cutoff to model the continuum and local absorption. We are able to confirm anti-correlations between the photon index and the luminosity and, for low fluxes, between the folding energy and the luminosity, implying a change of properties in the Comptonising plasma. 
We were not able to confirm a previously seen correlation between the cyclotron line energy and the luminosity of the source in the overall observation, but we observed a drop in the cyclotron line energy following a strong flare.
We see strong variability in absorption between the two observations and within one observation (for the $\sim$0.36--0.52 orbital phases) that can be explained by the presence of a large-scale structure, such as accretion and photoionisation wakes in the system, and our variable line of sight through this structure.
}


   \maketitle

\section{Introduction}
\label{section:intro}

\object{Vela X-1} is an eclipsing high-mass X-ray binary (HMXB) that consists of a B0.5 Ib supergiant, \object{HD 77581} \citep{Hiltner_1972}, and an accreting neutron star with a pulse period of $\sim$283\,s \citep{McClintock_1976} in an $\sim$8.964\,d orbit around this supergiant \citep{Kreykenbohm_2008}. A thorough review of the Vela X-1 system is given by \citet{Kretschmar_2021a}, who also discussed the role of this system in understanding HMXB systems in general. Here, we limit ourselves to points that are of immediate relevance for this work. The radius of HD~77581 is 30\,$R_{\odot}$ and the orbital separation $\sim$1.7\,$R_{\star}$ \citep{vanKerkwijk_1995, Quaintrell_2003}. The neutron star is thus embedded in the dense wind of the supergiant companion, which has a mass loss rate of $\sim$10$^{-6}M_{\odot}\,$yr$^{-1}$ \citep{Watanabe_2006}. The source is located at $1.99^{+0.13}_{-0.11}$ kpc \citep{Kretschmar_2021a} and is therefore one of the brightest persistent point sources in the X-ray sky despite a moderate mean luminosity of $5\times10^{36}$\,erg\,s$^{-1}$ \citep{Fuerst_2010a}. The mass of the neutron star is estimated to be $\sim$1.7--2.1 $M_{\odot}$ \citep{Kretschmar_2021a}.

The system is highly inclined \citep[>73\degree;][]{vanKerkwijk_1995}, which facilitates the study of the accretion and wind physical properties through observations at different orbital phases \citep[e.g.][]{Haberl_1990, Goldstein_2004, Watanabe_2006, Fuerst_2010a}. \citet{Doroshenko_2013a} showed a systematic change in the absorption along the orbit when averaged among multiple orbits: in particular, average absorption decreased after the eclipse, reached a minimum at the orbital phase of $\phi_{\mathrm{orb}} \approx 0.3, $ and grew afterwards, with a strong increase around $\phi_{\mathrm{orb}} \approx 0.5$. However, measurements of absorption at the same orbital phase are often different during different orbits of the system \citep{Kretschmar_2021a}. \citet{Fuerst_2010a} also found deviations from a log-normal distribution in the histogram of the orbital phase averaged brightness distribution of Vela X-1 that could be due to the complex and turbulent accretion geometry of the system.

The overall large-scale wind structure in Vela X-1 has been investigated in different publications over the years. \citet{Odaka_2013} interpreted the strong changes in absorption of Vela X-1 on short timescales observed with \textsl{Suzaku} as being due to the compact object's motion in the supersonic stellar wind forming a bow shock. \citet{Eadie_1975} suggested an accretion wake in Vela X-1 based on the absorption dips that appeared in the light curve of the \textsl{Ariel V} Sky Survey Experiment. Accretion and photoionisation wakes, and possibly a tidal stream, have also been suggested as causes \citep[see e.g.][]{Nagase_1983, Sato_1986, Blondin_1990a, Kaper_1994a, vanLoon_2001, Malacaria_2016a}.

An accretion wake forms through the focussing of the stellar wind due to the influence of the gravitational field of the neutron star, leading to an unsteady bow shock in the vicinity of the neutron star \citep{Blondin_1991, Manousakis_2015, Malacaria_2016a}. A photoionisation wake is formed when the wind interacts with the Strömgen sphere, which is created around the neutron star because of the photoionisation of the wind material \citep[see Fig.~2 in][]{Watanabe_2006}.

Hydrodynamical simulations optimised for Vela X-1 have been conducted in \citet{Blondin_1990a} and \citet{Manousakis:PhD}, who showed the formation of the wakes and an overall complex large-scale structure in the wind. \citet{Grinberg_2017} illustrated the accretion and photoionisation wakes in the Vela X-1 system in their Fig.~1 based on the simulations published in \citet{Manousakis:PhD}.

Even beyond the variable absorption, Vela X-1 is known to be an intrinsically highly variable source.\ It has shown bright flares \citep{Martinez_2014,Kreykenbohm_2008,Lomaeva_2020a} and off-states \citep{Kreykenbohm_2008, Doroshenko_2011a} where the observed flux decreased to less than 10\% of its normal value \citep{Kreykenbohm_2008,Sidoli_2015a}. 

The off-states have been interpreted in the context of a highly structured wind of the companion by some authors \citep{Kreykenbohm_2008, Ducci_2009, Fuerst_2010a}. For \citet{Manousakis_2015}, hydrodynamic instabilities are sufficient to explain the origin of the off-states without the need for intrinsic inhomogeneities in the stellar wind. Another theory to explain the origin of the off-states in Vela X-1 is the interaction of the neutron star's magnetosphere with the plasma \citep{Doroshenko_2011a}. Beyond the off-states, variations in the mass-accretion rate that lead to the observed flares and the variable absorption require, at least to some degree, a clumpy structure of the wind \citep{Kreykenbohm_2008, Fuerst_2010a, Martinez_2014}.

Vela X-1 shows various features in its X-ray spectrum that are typical for accreting neutron stars. One such feature is a prominent  fluorescent emission line associated with $\rm{FeK\alpha}$  around 6.4\,keV, which was first reported in \citet{Becker_1978}. Some previous analyses with high resolution instruments \citep[e.g.][]{Watanabe_2006} have also detected $\rm{FeK\beta}$, while others, at different orbital phases, struggled to constrain this feature \citep{Amato_2021a}. \citet{Watanabe_2006} stated there are three possible explanations for the presence of these lines in Vela X-1: the extended stellar wind, reflection off the stellar photosphere, and the accretion wake. Another observed line-like feature in the spectrum of X-ray pulsars is the so-called 10\,keV feature. The physical origin of this feature is still unknown \citep{Coburn_2002}, which may be due to our incomplete understanding of the continuum spectrum; it is usually modelled with simple phenomenological models. It appears to be an inherent feature in the spectra of accreting pulsars \citep{Coburn_2002} and probably reflects the limitations of the simple phenomenological models used \citep{Staubert_2019}. 
It may appear in emission in some sources, such as 4U 0115+63 \citep{Ferrigno_2009a, Mueller_2012} and EXO 2030+375 \citep{Klochkov_2007}. It is then modelled with a broad Gaussian emission component and is often referred to as the `10\,keV bump' model \citep{Reig_2013}. In other sources, it has been found in absorption, such as in XTE J0658–073 \citep{McBride_2006, Nespoli_2012}, Cen X-3 \citep{Santangelo_1998},  or Vela X-1 \citep{Fuerst_2014a}, and modelled with a broad Gaussian absorption component \citep{Fuerst_2014a}. In 4U 1901+03 \citep{Reig_2016a} and KS 1947+300 \citep{Fuerst_2014b}, similar features have been interpreted as being of possible magnetic origin.

Cyclotron resonant scattering features (CRSFs, or cyclotron lines) appear as broad absorption lines in the X-ray spectra of highly magnetised neutron stars whose magnetic field strength can be then directly measured from the CRSF energy. The fundamental CRSF is the result of resonant scattering of photons by electrons in strong magnetic fields from the ground level to the first excited Landau level followed by radiative decay \citep[see][and references therein, for a review]{Staubert_2019}. At higher excited Landau levels, the resulting CRSFs are called harmonics. The spectrum of Vela X-1 has been confirmed to show two CRSFs \citep{Kreykenbohm_2002}: a prominent harmonic line at $\sim$55\,keV and a weaker fundamental at $\sim$25\,keV \citep{Kendziorra_1992,Kretschmar_1997,Orlandini_1998,Kreykenbohm_1999a,Kreykenbohm_2002, Fuerst_2014a}. In Vela X-1, the harmonic line is broader and deeper than the fundamental line, which can be so weak that it cannot be significantly detected in some observations \citep{Odaka_2013}. The study of the CRSF energy variability with luminosity is important for determining the physical properties of the source (see Sect. \ref{section:accretion_regime}).

In several accreting X-ray pulsars, it was observed that the correlation of the CRSF energy with the luminosity and the correlation of the photon index with luminosity were inverse to each other, that is, if one was positive, the other was negative \citep[e.g.][for a sample study]{Klochkov_2011}. There are theoretical expectations for a correlation between the CRSF energy and luminosity. 

Following \citet{Becker_2012}, the variation in the CRSF energy with luminosity is due to the variation in the emission height within the accretion column that characterises the line-forming region. They define a source as being in the supercritical accretion regime when the radiation field has a dynamic effect on the infalling plasma. In this regime, the accretion flow is decelerated in the extended radiative shock, the height of which increases with the mass-accretion rate. It can move the line-forming region away from the surface of the neutron star to the higher altitudes corresponding to the lower magnetic field, though not necessarily up to the shock height \citep[see e.g.][]{Nishimura_2014}. This can explain the negative correlation of the CRSF centroid energy with luminosity. An alternative model has been suggested by \citet{Poutanen_2013} based on the reflection of the radiation from the neutron star surface at different altitudes. The so-called critical luminosity, $L_\mathrm{crit}$, is associated with the transition from the negative correlation of the CRSF energy and luminosity to the positive one. It is understood that in the subcritical regime (the X-ray luminosity $L_X<L_\mathrm{crit}$), the role of the radiation in the stopping of the accretion flow becomes negligible. There are several mechanisms proposed to explain the matter deceleration in the subcritical regime that can also explain the observed positive correlation of the CRSF energy with the luminosity. The matter can be fully decelerated (i.e. stopped) by Coulomb collisions in the accretion channel. The height at which the effective stopping occurs decreases with the increasing luminosity \citep{Staubert_2007,Becker_2012}, causing the positive correlation. Another scenario \citep{Mushtukov_2015b} suggests that the shift in the CRSF energy to lower values with the decreasing luminosity can also be explained by the Doppler effect in the accretion channel moderated by the luminosity-dependent velocity profile. Another possibility at low mass-accretion rates is the origination of the collisionless shock above the surface \citep{Langer_1982}. The height of this shock decreases with increasing mass-accretion rate \citep{Shapiro_1975}, thus moving the line-forming region closer to the surface of the neutron star and resulting in the positive correlation of the CRSF energy and luminosity \citep{Rothschild_2017,Vybornov_2017}.

The value $L_\mathrm{crit}$, corresponding to the transition between the subcritical and supercritical regimes, is dependent on the magnetic field of the source and the geometry of the accretion channel. The analytical expressions were obtained, for example, by \citet{Basko_1976} and \citet{Becker_2012} under different considerations. The critical luminosity was calculated numerically as a function of the magnetic field of the neutron star for different types of accretion (disc or wind) by  \citet{Mushtukov_2015a}, taking the resonance in the Compton scattering cross-sections and the possible mixture of polarisation modes into account. It was shown that the critical luminosity is not a monotonic function of the magnetic field strength. 

For Vela X-1, a tentative detection of such a correlation between CRSF energy and luminosity has been presented in \citet{Fuerst_2014a}. It was also found that $\Gamma$ was anti-correlated with the luminosity \citep[e.g.][]{Odaka_2013,Fuerst_2014a}, indicating that the source was in a subcritical accretion regime, behaving similarly to sources such as Her X-1 \citep{Staubert_2007} and GX 304--1 \citep{Yamamoto_2011}. 

Given the importance of Vela X-1 for the study of wind-accreting neutron stars, we use two further observations with \nustar, taken at different orbital phases of the source, to map the accretion environment and the structure of the stellar wind in the source and to further investigate possible correlations of the CRSF with other spectral parameters. First, we describe the two datasets and the analysis software used in Sect.~\ref{section:observation_reduction}. We then present the light curves and timing results of those observations in Sect.~\ref{section:lightcurves}. We proceed to time-averaged spectroscopy in Sect.~\ref{section:average_spectro} and to time-resolved spectroscopy in Sect.~\ref{section:timeres_spectro}. We discuss the results in Sect.~\ref{section:discussion}, focussing on the wind structure and the CRSFs, and give a summary and an outlook in Sect.~\ref{section:summary_outlook}.

\section{Observation and data reduction}
\label{section:observation_reduction}

\begin{figure}
    \centering
    \centerline{\includegraphics[trim=0cm 0cm 0cm 0cm, clip=true, width=1.0\linewidth]{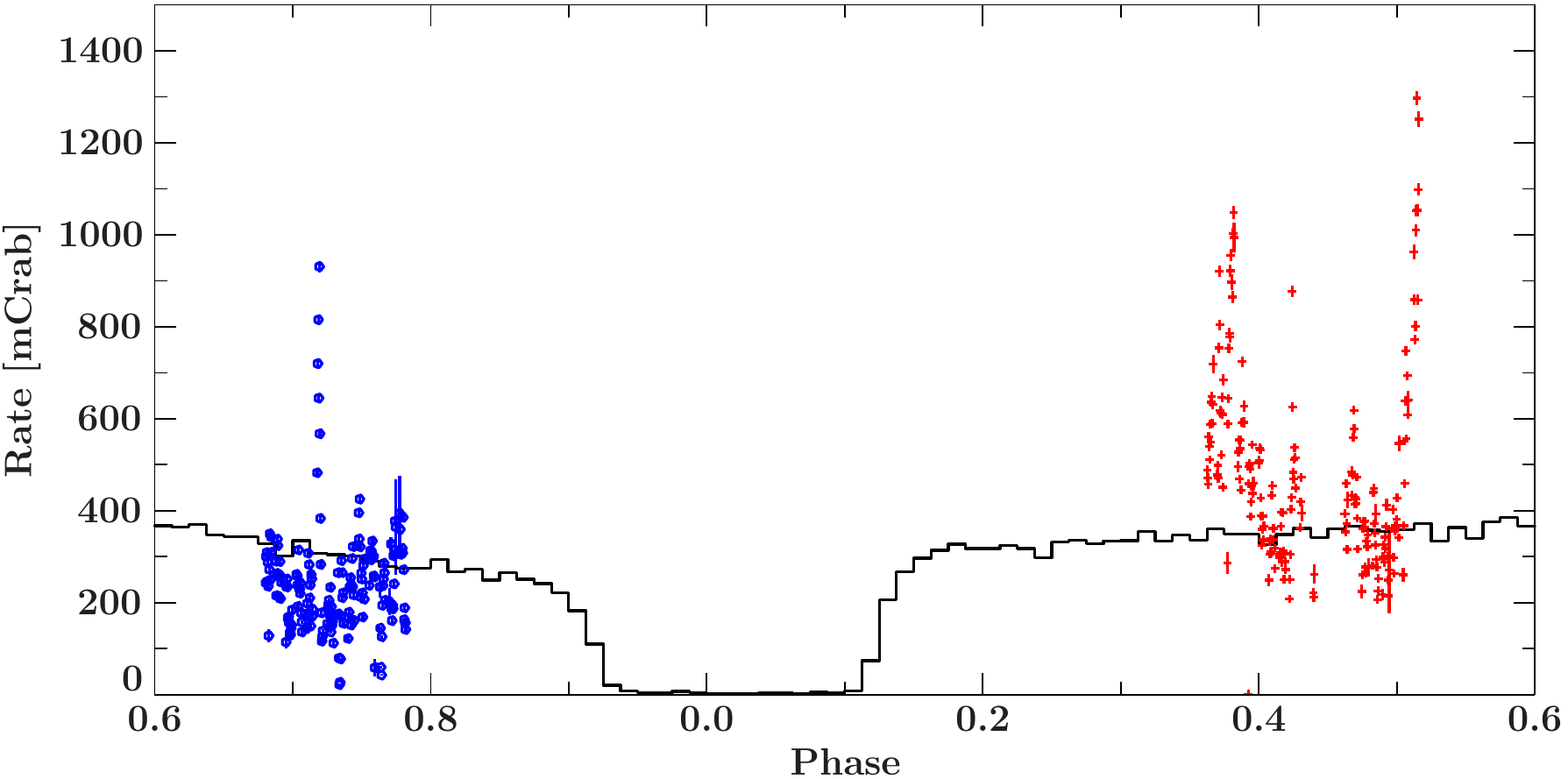}}
    \caption{\nustar 3--79\,keV flux for observation I (blue) in January 2019 and observation II in May 2019 (red) with a time resolution of $P_{\rm{I}} = 283.4532 \, \rm{sec}$ and $P_{\rm{II}} = 283.4447 \, \rm{sec,}$ respectively, and \textsl{Swift}/BAT 15--50\,keV flux averaged over all data since 2005 (solid black line) plotted over the neutron star's orbital phase, where $\phi_{\rm{orb}}=0$ is defined with $T_{90}$.}
    \label{fig:bat_nustar_orbitphase}
\end{figure}

\nustar \citep{NuSTAR_team_2013} observed Vela X-1 on 10--11~January~2019 and on 3--5~May~2019 as science target using the focal plane modules A and B (FPMA and FPMB). We then have two datasets with an exposure of $\sim$36\,ks and $\sim$40\,ks, referred as observation I and observation II, respectively. Details about the observations are given in Table \ref{tab:obs_details} and Fig.~\ref{fig:bat_nustar_orbitphase} shows the count rate of the Vela X-1 system against the orbital phase. As can be seen from Fig.~\ref{fig:bat_nustar_orbitphase}, observation II consists of two parts. The interruption is due to data loss caused by problems with the ground station and the corresponding data cannot be recovered. 
This sketch also includes the \textsl{Swift}/BAT 15--50\,keV light curve averaged over all data since 2005 and therefore representing the averaged behaviour of the source smoothing out the variability on shorter timescales \citep{Fuerst_2010a,Kretschmar_2021a}.
The eclipse is clearly visible around $\phi_{\rm{orb}}=0$.  

\begin{table*}
\renewcommand{\arraystretch}{1.1}
\caption{Observation details.}
\label{tab:obs_details}
\begin{center}
\begin{small}
\begin{tabular}{llccccc}    
\hline\hline
Name &
Obs ID & 
\multicolumn{1}{c}{Time Start} & 
\multicolumn{1}{c}{Time Stop} &
\multicolumn{1}{c}{Exposure} &
\multicolumn{1}{c}{Orbital phase} &
\multicolumn{1}{c}{Orbital phase} \\
& &
\multicolumn{1}{c}{MJD (day) binarycor} & 
\multicolumn{1}{c}{MJD (day) binarycor} &
\multicolumn{1}{c}{(ks)} &
\multicolumn{1}{c}{(with $T_{\mathrm{90}}$)} &
\multicolumn{1}{c}{(with $T_{\mathrm{ecl}}$)} \\
\hline
\multicolumn{1}{l}{Obs I} & 90402339002     & 58493.1813     & 58494.0910     & 36.086       & 0.68--0.78     & 0.65--0.75       \\
Obs II & 30501003002     & 58606.8688     & 58608.2465     & 40.562       & 0.36--0.52     & 0.34--0.49       \\
\textit{Obs IIa} & \textit{30501003002}     & \textit{58606.8688}     & \textit{58607.5660}     & \textit{22.557}       & \textit{0.36--0.44}     & \textit{0.34--0.42}       \\
\textit{Obs IIb} & \textit{30501003002}     & \textit{58607.7625}     & \textit{58608.2465}     & \textit{18.005}       & \textit{0.46--0.52}     & \textit{0.44--0.49}       \\
\hline
\multicolumn{6}{p{0.7\linewidth}}{As observation II is divided into two parts, observations IIa and IIb, for the analysis, we list the details for both sub-observations above.}
\end{tabular}
\end{small}
\end{center}
\renewcommand{\arraystretch}{1.0}
\end{table*}

The orbital phases $\phi_\mathrm{orb}$ are derived from the ephemeris from \citet{Kreykenbohm_2008} and \citet{Bildsten_1997} (see Table~\ref{tab:ephemeris}). Several definitions of the time of phase zero can be found in the literature. The most common are $T_{\rm{ecl}}$ and $T_{90}$. $T_{\rm{ecl}}$ is the mid-eclipse time whereas $T_{90}$ is the time when the mean longitude $l$ is equal to 90\degree. For the particular case of Vela X-1, those two points are very close to each other, the difference is of the order of $\Delta \phi \approx 0.02$.  Nonetheless, in soft X-rays the eclipse boundaries can be hard to define and considering that the orbit is elliptical, defining $T_{\rm{ecl}}$ is not always that clean. Explanations on how to convert $T_{90}$ to $T_{\rm{ecl}}$ can be found in \citet{Kreykenbohm_2008}. In this work, we exclusively use $T_{90}$. 

\begin{table*}[!htpb]
\renewcommand{\arraystretch}{1.1}
\caption{Ephemeris used.}
\label{tab:ephemeris}
\begin{center}
\begin{small}
\begin{tabular}{llrl}    
\hline\hline
Reference & 
Orbital parameter &
\multicolumn{1}{c}{Value} &
\multicolumn{1}{c}{Units} \\
\hline
\citet{Kreykenbohm_2008}     & Time of mean longitude equal to $\pi/2$ ($T_{\mathrm{90}}$)     & 52974.001 $\pm$ 0.012     & MJD (day) \\
\citet{Kreykenbohm_2008}     & Mid-eclipse time ($T_{\mathrm{ecl}})$                             & 52974.227 $\pm$ 0.007     & MJD (day) \\
\citet{Kreykenbohm_2008}     & Orbital period ($P_{\mathrm{orb}}$)                             & 8.964357 $\pm$ 0.000029  & day \\
\citet{Bildsten_1997}     & Semi-major axis ($a \sin i$)                                     & 113.89 $\pm$ 0.13      & lt-sec \\
\citet{Bildsten_1997}     & Eccentricity ($e$)                                              & 0.0898 $\pm$ 0.0012    & \\
\citet{Bildsten_1997}     & Longitude of periastron ($\omega$)                                  & 152.59 $\pm$ 0.92      & \\
\citet{vanKerkwijk_1995}     & Inclination ($i$)                                               & $> 73$                   & \degree\\
\hline
\end{tabular}
\end{small}
\end{center}
\renewcommand{\arraystretch}{1.0}
\end{table*}

The two observations are carried out at different orbital phases. At the orbital phase of observation I that is at $\phi_\mathrm{orb} \approx 0.75$, the accretion and photoionisation wakes are placed along the line of sight of the observer (see Fig.~1 in  \citealt{Grinberg_2017}). Thus, a different behaviour of the absorption is expected compared to observation II at $\phi_\mathrm{orb} \approx 0.4$--0.5, where the wakes may be starting to pass through the observer's line of sight.

We use the NUSTARDAS pipeline v2.0.0 and HEASOFT v6.28 with \nustar CALDB (calibration database) v20200826 applied with the clock correction to extract spectra and light curves for the time-averaged observations and we also proceed to extract spectra orbit-by-orbit and pulse-by-pulse, necessary for a time-resolved analysis of this highly variable source (see Sect.~\ref{section:timeres_spectro}). For the orbit-by-orbit spectroscopy, we extract a spectrum for each NuSTAR orbit around the Earth. That leads to 14 spectra for observation I and 19 spectra for observation II. For the pulse-by-pulse spectroscopy, we extract a spectrum for each rotation of the neutron star, with the corresponding integration pulse period derived for this observation (see Sect.~\ref{section:lightcurves}). That leads to 168 spectra for observation I and 209 spectra for observation II. The event times are barycentred using the \texttt{barycorr} tool from NUSTARDAS pipeline and corrected for the binary orbit using the ephemeris from Table \ref{tab:ephemeris}. We use the Interactive Spectral Interpretation System (ISIS) v1.6.2-47 \citep{Houck&Denicola_2000} to analyse the data and note that ISIS allows access to XSPEC \citep{Arnaud_1996a} models that are referenced later in the text.

We extracted source spectra from a region with a radius of $\sim$82 arcsec  for observation I and $\sim$67 arcsec for observation II around Vela X-1’s FK5 coordinates, separately for FPMA and FPMB. As Vela X-1 is very bright, it illuminates the whole focal plane, and we thus extract background spectra circular regions with $\sim$63 arcsec radius for observation I and $\sim$67 arcsec radius for observation II as far away from the source as possible for both observations to minimise source influence on background estimation.
Since the background changes over the field-of-view of the instrument, systematic uncertainties are formally introduced by this method. Vela X-1, however, is about a factor  of 5 brighter than the background even at the highest energies used here, such that the effect of residual uncertainties is negligible.

Due to a thermal blanket tear, the detector focal plane module FPMA requires a low energy effective area correction \citep{Madsen_2020}. Such a correction is automatically applied in all \nustar CALDB releases starting with the 20200429 CALDB. However, for some sources the automatic procedure results in an over-correction. This is the case for our observation II, where the automatic procedure results in stark differences between the focal plane modules. After consulting with the \nustar calibration team (K. Madsen, priv. comm), it was decided that the best approach was to revert to the old FPMA ARF\footnote{\url{https://nustarsoc.caltech.edu/NuSTAR_Public/NuSTAROperationSite/mli.php}} for this observation, which results in an agreement between the modules.

All spectra were re-binned within ISIS to a minimal signal to noise of 5, adding at least 2, 3, 5, 8, 16, 18, 48, 72, and 48 channels for energies between 3.0–10, 10–15, 15–20, 20–35, 35-45, 45-55, 55-65, 65-76, and 76-79\,keV, respectively. Uncertainties are given at the 90\% confidence ($\Delta \chi^2 = 2.7$ for one parameter of interest), unless otherwise noted.

\section{Light curves and timing}
\label{section:lightcurves}

\subsection{Pulse period}

The pulse period of Vela X-1 shows strong variations on all timescales mainly due to a highly variable accretion rate. It varies in a way mostly consistent with random-walk \citep{deKool&Anzer_1993}. To measure this period accurately to do the pulse-by-pulse analysis in Sect.~\ref{section:timeres_spectro}, we perform epoch-folding \citep{Leahy_1987} on the FPMA extracted light curve with 1 sec binning. A pulse period of $P_{\rm{I}} = 283.4532 \pm 0.0008\,\rm{sec}$ for observation I and of $P_{\rm{II}} = 283.4447 \pm 0.0004\,\rm{sec}$ for observation II are found, which is consistent with the overall pulse period history of Vela X-1 from \textsl{Fermi} Gamma-ray Burst Monitor\footnote{\url{https://gammaray.nsstc.nasa.gov/gbm/science/pulsars/lightcurves/velax1.html}}. The pulse period changes by such a small fraction ($+/- \sim$0.06$\%$ at most) within one observation that it is not enough to induce significant shifts in the pulse phase in the pulse-by-pulse analysis. The uncertainties were estimated using a Monte Carlo simulation \citep{Larsson_1996} with 2000 runs.

\subsection{Light curves}

\begin{figure*}[!htpb]
    \centering
    \includegraphics[trim=0cm 0cm 0cm 0.5cm, clip=true, width=0.48\linewidth]{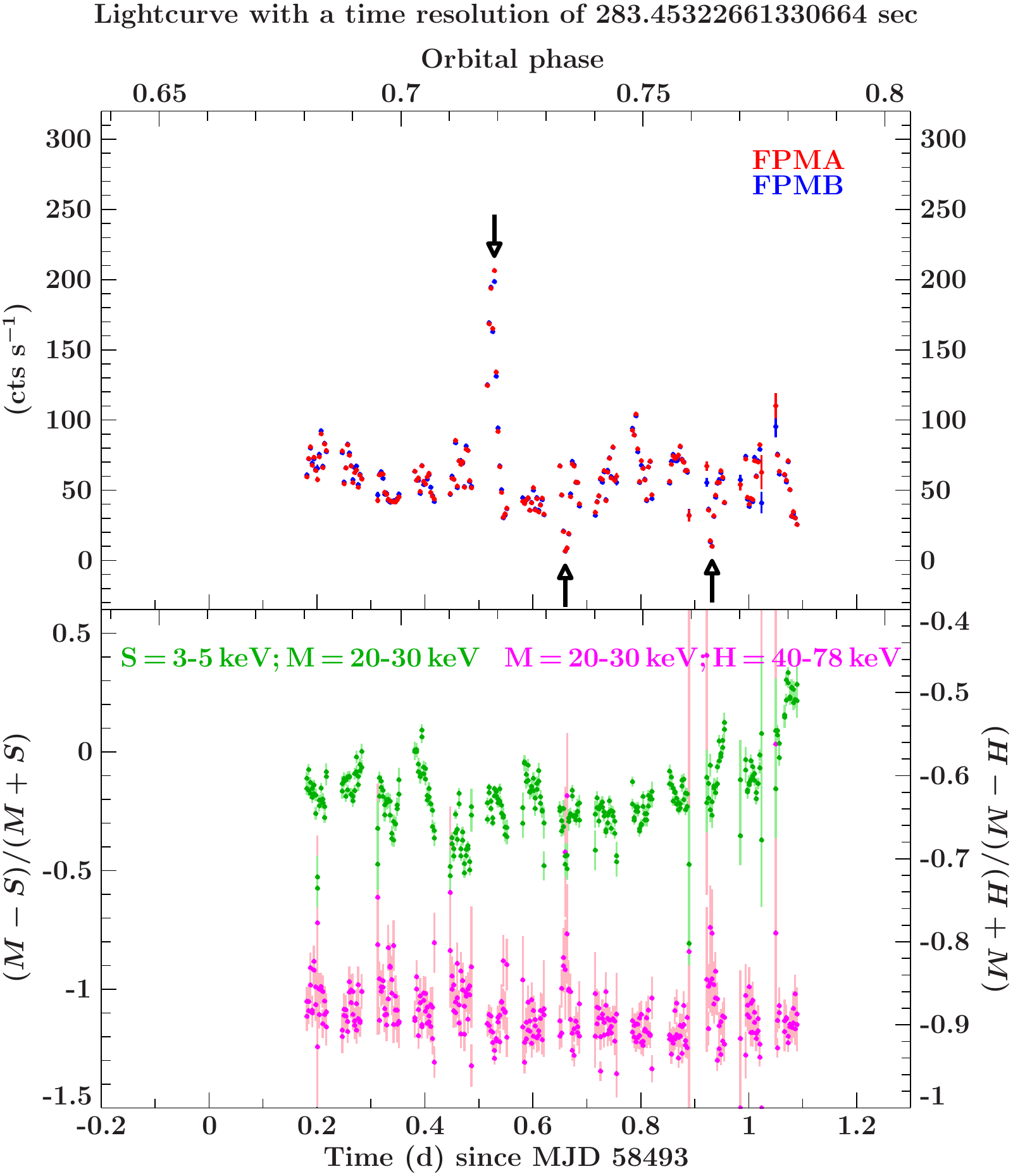}\hfill
    \includegraphics[trim=0cm 0cm 0cm 0.5cm, clip=true, width=0.48\linewidth]{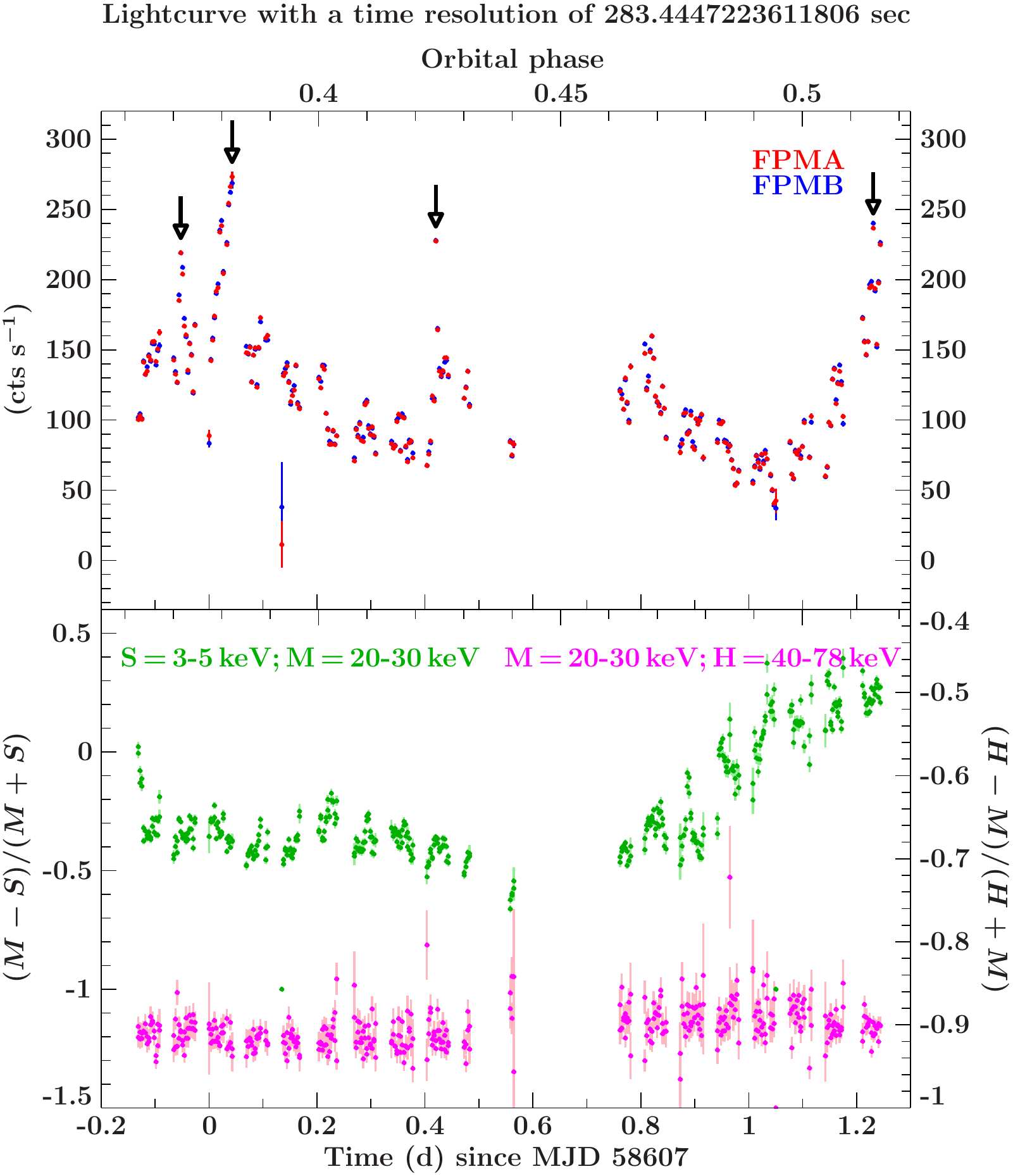}
    \caption{Light curves and hardness ratios for observations I (\textsl{left panel}) and II (\textsl{right panel}) with a time resolution of
    $P_{\rm{I}} = 283.4532 \, \rm{sec}$ and $P_{\rm{II}} = 283.4447 \, \rm{sec,}$ respectively. Major off-states and major flares are indicated by arrows. Green points show the hardness ratio between the 3-5\,keV and the 20-30\,keV bands using the left-hand $y$ axis. Magenta points show the hardness ratio between the 20-30\,keV and the 40-78\,keV bands using the right-hand $y$ axis.}
    \label{fig:hr_lc}
\end{figure*}

Vela X-1 is known to be a highly variable source, and thus we present the light curves of observations I and II in the top panels of Fig.~\ref{fig:hr_lc} to investigate the relevant features. We bin the light curves to the spin period of the neutron star to avoid the intra-pulse variability.

In observation I, at $T_{\rm{obs}} \approx 58493.53 \, \rm{MJD}$, we can observe a large flare indicated in the left upper panel of Fig.~\ref{fig:hr_lc} where the source flux increases to $\sim$200\,counts$\,$s$^{-1}$ per module, from an average of $\sim$60\,counts$\,\rm{s}^{-1}$ during the whole observation. Moreover, two off-states are detected at $T_{\rm{obs}} \approx 58493.66 \, \rm{MJD}$ and $T_{\rm{obs}} \approx 58493.93 \, \rm{MJD}$. Here, the source flux drops to $\sim$6-7\,counts$\,\rm{s}^{-1}$ for the first off-state and $\sim$10\,counts$\,\rm{s}^{-1}$ for the second one.

In observation II, the average source flux is $\sim$118\,counts$\, \rm{s}^{-1}$ about twice as much as during observation I. At $T_{\rm{obs}} \approx 58606.95 \ \rm{MJD}$, $T_{\rm{obs}} \approx 58607.03 \, \rm{MJD}$, $T_{\rm{obs}} \approx 58607.42 \, \rm{MJD}$ and $T_{\rm{obs}} \approx 58608.23 \, \rm{MJD}$ four major flares indicated in the right upper panel of Fig.~\ref{fig:hr_lc} are detected reaching $\sim$220\,counts$\,\rm{s}^{-1}$, $\sim$275\,counts$\,\rm{s}^{-1}$, $\sim$228\,counts$\,\rm{s}^{-1}$ and $\sim$238\,counts$\,\rm{s}^{-1}$, respectively. However, this time no off-state is detected. The big gap in observation II between $T_{\rm{obs}} \approx 58607.57 \, \rm{MJD}$ and $T_{\rm{obs}} \approx 58607.76 \, \rm{MJD}$ is due to the loss of data.

\subsection{Hardness ratios}

In the bottom panels of Fig.~\ref{fig:hr_lc}, we present the hardness ratios calculated as $HR = (\rm{Medium}-\rm{Soft})/(\rm{Medium}+\rm{Soft})$ (left-hand y axis) and $HR = (\rm{Hard}-\rm{Medium})/(\rm{Hard}+\rm{Medium})$ (right-hand y axis) with three different energy bands. The low-energy ratio (in green) covers the energy region where most of the absorption from the stellar wind happens, and thus, we tried to compare it to the high-energy ratio (in magenta), which is less impacted by the effect of the stellar wind absorption and instead more representative of the true underlying continuum shape.

In observation II, we observe that the 3--5\,keV to 20--30\,keV hardness behaves differently before and after the data loss period: it is roughly constant before and rising after, while the 20--30\,keV to 40-78\,keV hardness shows no such trends. This motivated us to divide observation II into two parts to be analysed separately: observation IIa and observation IIb, which correspond to the data before and after the loss, respectively (details can be found in Table \ref{tab:obs_details}). Moreover, the time-averaged modelling of the spectrum of observation II leads to a bad description of the data, supporting the choice of a separate analysis (see Sect.~\ref{section:average_spectro}).

In observation I, there are two changes, at $T_{\rm{obs}} \approx 58493.67 \, \rm{MJD}$ and $T_{\rm{obs}} \approx 58493.92 \, \rm{MJD,}$ that correspond to the location of the two off-states. However, there is no change associated with the flare in any of both hardness ratios. This has already been observed in the flares analysed in \citet{Kreykenbohm_2008} or \citet{Fuerst_2014a}. The spectrum becomes slightly harder in the low-energy ratio towards the end of the observation, but the hardness shows no other strong overall trends.

In observation II, the high-energy hardness ratio shows the exact same behaviour than during observation I. However, this time we do not see any strong change as no off-state was observed during this observation. Again, no significant change in both hardness ratios is shown at the location of the flares. The spectrum in the low-energy ratio remains roughly flat for the first part of the observation but becomes suddenly harder starting from $T_{\rm{obs}} \approx 58607.75 \, \rm{MJD}$, which corresponds to the end of the period where the data were lost.

The changes in the hardness ratios observed during the off-states and the time variability of Vela X-1 during observations I and II require a spectral analysis of the source on shorter timescales. To achieve this goal, we start by analysing the spectra averaged over the three effective observations (Obs~I, Obs~IIa, and Obs~IIb) as a basis for the further analysis (Sect.~\ref{section:average_spectro}).
We then continue analysing the orbit-by-orbit spectra  to have a first overview of the evolution of various spectral parameters over time and use these values as inputs for the pulse-by-pulse spectroscopy. Both analyses are detailed in Sect.~\ref{section:timeres_spectro}. This step-by-step time-resolved spectral analysis allows us to have a fine adjustment of the parameters and a follow-up of the parameters along the time at each step of the analysis. Finally, the obtained results are discussed in Sect.~\ref{section:discussion}.

\section{Time-averaged spectroscopy}
\label{section:average_spectro}

\subsection{Choice of continuum model}

We first address the choice of the overall continuum model. Different flavours of power law models with high energy cutoffs are used in the literature to empirically describe the continuum spectral shape of accreting neutron stars \citep[e.g.][]{Staubert_2019}.

The Fermi-Dirac cutoff \citep[\texttt{FDcut;}][]{Tanaka_1986a} is the most widely used continuum model for Vela X-1 (see \citealt{Kreykenbohm_1999a,Kreykenbohm_2008,Fuerst_2014a}). Thus, we used a power law with the photon index $\Gamma$ and \texttt{FDcut} with the cutoff energy $E_{\rm{cut}}$ and the folding energy $E_{\rm{fold}}$ so that

\begin{align}
\label{eq:fdcut}
F(E) \propto E^{-\Gamma} \times \left(1+ \exp \left(\frac{E-E_{\rm{cut}}}{E_{\rm{fold}}} \right) \right) ^{-1}.
\end{align}
In this model, we constrain $E_{\rm{fold}}$ and $E_{\rm{cut}}$ to 4--18\,keV and 18--40\,keV, respectively, to avoid model degeneracies. With this model, we are able to achieve a good description of the data (see Sect.~\ref{section:Modelling} for an in-depth discussion).

We also investigated several other models to fit all our observations. The same trends have been observed for all of them, but for simplicity we only quote the values for observation I below. The best-fit parameters for the time-averaged tested models can be found in the appendix (Tables~\ref{fig:app1} and \ref{fig:app2}).

We attempt to describe the data with a \texttt{highecut}\footnote{\url{https://heasarc.gsfc.nasa.gov/xanadu/xspec/manual/node129.html}\label{footnote:xspec_url}} model, another often used model for the continuum in accreting neutron stars \citep[]{Santangelo_1998, Staubert_2019}. It results in a statistically worse description of the data than the \texttt{FDcut} model. 

We also used the \texttt{NPEX} \citep{Mihara_1995} model for the continuum. This model consists of the sum of a negative and a positive power law, which is then modified by an exponential cutoff that is characterised by a folding energy. This model has previously been used for a number of HMXBs \citep[see e.g.][]{Hemphill_2014,Jaisawal_2016}. We explored the \texttt{NPEX} with the 10\,keV feature parameter space (see Sect.~\ref{section:Modelling} for a more detailed discussion about the modelling of the 10\,keV feature). The combination of two power laws with wide allowed ranges of indices makes the analysis difficult and provides several combinations of good fits. If the positive power law is fixed to 2, the function is known to be a good approximation of the
unsaturated Comptonisation spectrum in neutron stars \citep{Makishima_1999,Odaka_2013,Hemphill_2014}. This results in a good fit ($\chi^2/{\rm{dof}} \approx 604.15/456$), with both power law components of \texttt{NPEX} contributing to the continuum, but both CRSFs are very broad and deep (see Sect.~\ref{section:Modelling} for a more detailed discussion about the modelling of the CRSFs), in particular the strength of the harmonic CRSF reaches $83\pm10$, effectively altering the continuum. Moreover, the norm of the positive power law index was found close to 0, approaching a single power law with a high energy cutoff.  While statistically satisfactory, this model is thus not a good description of the data \citep[see e.g.][for discussion of similar problems for a different HMXB]{Bissinger_2020a}. We also explore the \texttt{NPEX} without the 10\,keV feature parameter space but it also results in a bad description of the continuum modelled by a too deep harmonic CRSF.

We further explore a Comptonisation continuum, using a single \texttt{compTT} \citep{Titarchuk_1994} in a spherical accretion case as done previously in \citet{Maitra&Paul_2013} for Vela X-1. This results in a statistically good description of the data for both with and without 10\,keV feature cases, but the overly strong CRSFs again effectively alter the continuum. A double \texttt{compTT} model has been recently proposed for several X-ray pulsars \citep[see e.g.][]{Doroshenko_2012,Tsygankov_2019}, albeit for the low luminosity regime below Vela X-1's luminosity. We also explore this double \texttt{compTT} model without the 10\,keV feature, as it should in theory get rid of the absorption feature observed in the 10\,keV range. We obtain a good fit ($\chi^2/{\rm{dof}}$ is $622.53/456$, slightly larger but similar to the \texttt{FDcut} best-fit) where the parameters can be constrained and the CRSFs behave as expected (if both $T_0$ are tied and allowed to vary up to 1\,keV, otherwise if they are free to vary in a wider range of values then the CRSFs become very strong again), showing around the same values as our \texttt{FDcut} best-fit model. However, the behaviour of the Comptonisation components does not agree with expectations from previous applications of this model and defies previous physical interpretation.
\citet{Tsygankov_2019}, who introduced this model for the low luminosity state of GX 304-1, discuss that the components are likely to be independent. In particular, the low-energy component ($kT \lesssim 2\,{\rm{keV}}$) can be explained as radiation from hotspots at the neutron star's surface heated up by the accretion process. However, in our analysis, this component is much hotter, with $kT \approx 6.9$\,keV. Additionally, in the low luminosity case, the two Comptonisation continua are of equal strength, while in our model the hotter component dominates the overall flux and spectral shape. 

Overall, we conclude that the \texttt{FDcut} model offers the best empirical description of the continuum that we can obtain.
Since it also allows us direct comparison with previous results, especially \citet{Fuerst_2014a} because different continua may lead to shifts in derived CRSF positions \citep[e.g.][]{Mueller_2012}, we decide to adopt it for further modelling.

\subsection{Modelling}
\label{section:Modelling}

We introduced a floating cross-normalisation parameter, $C_{\rm{FPMB}}$, in order to give the relative normalisation between \nustar's two detectors: FPMA and FPMB.

We try modelling the fluorescent emission lines features associated with $\rm{FeK\alpha}$ and $\rm{FeK\beta}$ with narrow Gaussian components. The $\rm{FeK\beta}$ line cannot be constrained because of the limited energy resolution of \nustar and the overlapping Fe K-edge at 7.1\,keV. Thus we only include the $\rm{FeK\alpha}$ line around 6.4\,keV in our final model.

For our analysis, we modelled the CRSFs using two multiplicative Gaussian absorption lines (\texttt{gabs} in XSPEC\footref{footnote:xspec_url}) corresponding to the fundamental CRSF (CRSF,F) and to the harmonic CRSF (CRSF,H) with
\begin{align}
    \rm{CRSF}(E) = \exp\left[-\left(\frac{d}{\sigma\sqrt{2\pi}}\right)\exp\left(-0.5\left(\frac{E-E_{cyc}}{\sigma}\right)^2\right)\right],
\end{align}
where $d$ is the line depth and $\sigma$ the line width. 

We described the 10\,keV feature with a broad Gaussian line in absorption, as also done in \citet{Fuerst_2014a}. We first tried to describe the spectrum without this feature. However, the analysis leads to visible residuals in absorption in the 10\,keV region for observation IIa and a bad description of the overall continuum, which is compensated for by deeper and broader CRSFs that effectively alter the continuum for all observations. We therefore conclude that the feature is necessary for a good description of our data. The physical origin of this 10\,keV feature is unclear. The 10\,keV feature in absorption has been found in other sources \citep{McBride_2006,Nespoli_2012,Santangelo_1998} and only sometimes tentatively associated with a CRSF \citep{Reig_2016a,Fuerst_2014b}. In Vela X-1, it has so far never been interpreted as a CRSF, although we cannot definitely exclude such an origin currently.

We further had to constrain the width of the fundamental line, setting it ad hoc to $\sigma_{\rm{CRSF,F}} = 0.5 \times \sigma_{\rm{CRSF,H}}$, following previous work \citep{Fuerst_2014a}. 
This value is based on the fact that the width of the CRSF is dominated by thermal broadening, for which $\sigma/E$ is independent of energy \citep{Schwarm_2017a}. As  $E_{\rm{CRSF,F}}/E_{\rm{CRSF,H}}$ is roughly 0.5, we used 0.5 for the width ratio as well. Letting the width free results in an overly wide fundamental CRSF that effectively describes the continuum in the 10--30\,keV range. We do, however, verify that when the line width is free, the best-fit CRSF energies, especially for the harmonic line, do not change significantly. Further, freezing the line width to a slightly larger value does not change the resulting best fit significantly. However, much narrower fundamental line (0.2 of the harmonic) leads to an increase in the reduced chi-square ($\chi^2/\rm{dof} \approx 716.96/456$ for observation I) explained by a too narrow fundamental line shifting down the line energy of $\sim$$1$\,keV.

To describe the different absorption components surrounding the neutron star, we use the \texttt{tbabs} model \citep{Wilms_2000} and the corresponding abundances and cross-sections \citep{Verner_1996}. We first try a single-absorption model but the spectrum is not well described at low energies, as can be seen on the example of observation I (Fig.~\ref{fig:step-by-step_Felix} c). This can be understood as a high contribution of the absorption from the stellar wind at low energies. Therefore, we use a partial covering model with the covering fraction $\rm{CF}$ to take into account the clumpy structure of the stellar wind. The allowed values for $\rm{CF}$ range between 0 and 1. Different setups of a partial covering model or equivalent models have previously been shown to be necessary to describe the spectrum of Vela X-1 well \citep[e.g.][]{Martinez_2014,Fuerst_2014a,Malacaria_2016a}.

Therefore, our final and best-fit model can be written as
\begin{align}
\label{eq:final_model}
\begin{split}
I(E) = &(\rm{CF} \times N_{\rm{H,1}} + (1-\rm{CF})) \times N_{\rm{H,2}} \\
& \times (F(E) \times \rm{CRSF,F} \times \rm{CRSF,H} + \rm{FeK\alpha} + 10\,\rm{keV}).
\end{split}
\end{align}

\begin{figure*}
    \centering
    \includegraphics[trim=0cm 0cm 0cm 15.1cm, clip=true, width=0.49\linewidth]{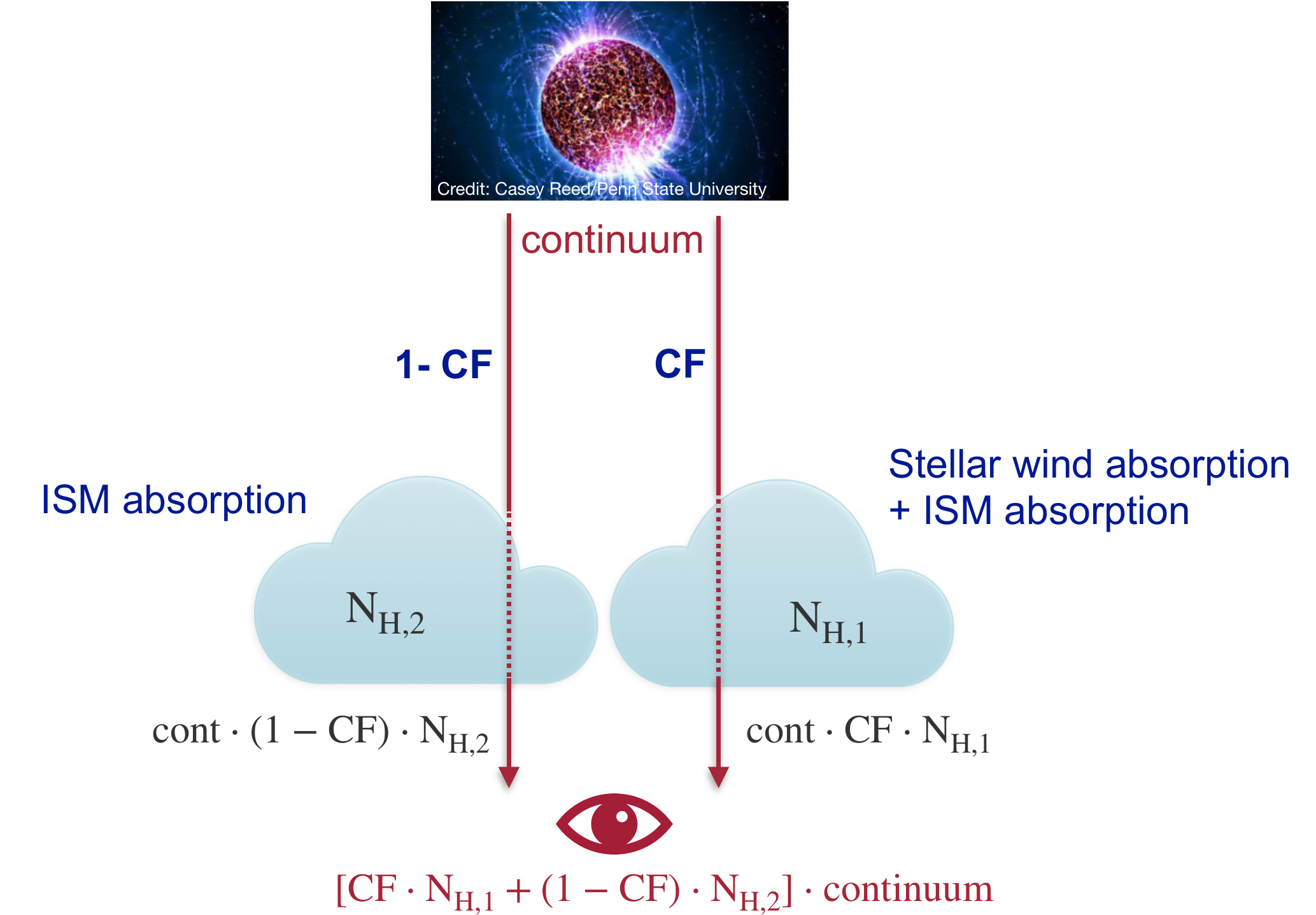} 
    \includegraphics[trim=0cm 15.1cm 0cm 0cm, clip=true, width=0.49\linewidth]{images/pc_model.pdf}
    \caption{Comparison of the partial covering model from \citet{Fuerst_2014a} (left side) and the partial covering model from this work (right side). The continuum corresponds to $(F(E) \times \rm{CRSF,F} \times \rm{CRSF,H} + \rm{FeK\alpha} + 10\,\rm{keV})$ in both cases.}
    \label{fig:pc_model}
\end{figure*}

This partial covering model is equivalent to the partial covering model used in \citet{Fuerst_2014a} and allows for an easier comparison of the spectral parameters. The only difference lies in the description of the absorption components (see Fig.~\ref{fig:pc_model}) that are differently parametrised between the two models. In \citet{Fuerst_2014a}, they split the absorption component into two absorption columns: one corresponding to the absorption from the interstellar medium and the other one corresponding to both the absorption from the interstellar medium and from the stellar wind. In this work, we used two distinct absorption columns to describe the stellar wind and the absorption from the interstellar medium in order to have access to both parameters separately.

The absorption $N_{\rm{H,1}}$ corresponding to the stellar wind embedding the neutron star is free and the absorption from the interstellar medium $N_{\rm{H,2}}$ has been fixed to $3.71 \times 10^{21}\,\rm{cm^{-2}}$ based on NASA's HEASARC $\rm{N_H}$ tool website\footnote{\url{https://heasarc.gsfc.nasa.gov/cgi-bin/Tools/w3nh/w3nh.pl}} \citep{HI4PI_2016}. We also tested to leave $N_{\rm{H,2}}$ as free parameter of the fit. For observation I, we obtain a high upper limit, consistent with the fixed value. For both parts of observation II, we obtain higher values that also lead to a systematic decrease in the photon index $\Gamma$. Given that \nustar does not cover the range below 3 \,keV, it is challenging to constrain low values of $N_{\rm{H},2}$, especially for a partial coverer. We thus decide to fix the $N_{\rm{H},2}$ to the theoretical galactic column density value for the following analysis. 

We first show the contribution from the broad line components in Figs.~\ref{fig:step-by-step_Felix} and~\ref{fig:step-by-step_Vic}. We observe a broad, strong contribution of the harmonic CRSF at around 54\,keV and a much weaker fundamental CRSF at $\sim$25\,keV (residual panel b), as commonly found for this source \citep{Kreykenbohm_2002,Fuerst_2014a}. There is also a visible contribution of the 10\,keV feature (residual panel c).

\begin{figure}[!htpb]
    \centering
    \centerline{\includegraphics[trim=0cm 0cm 0cm 0cm, clip=true, width=1.0\linewidth]{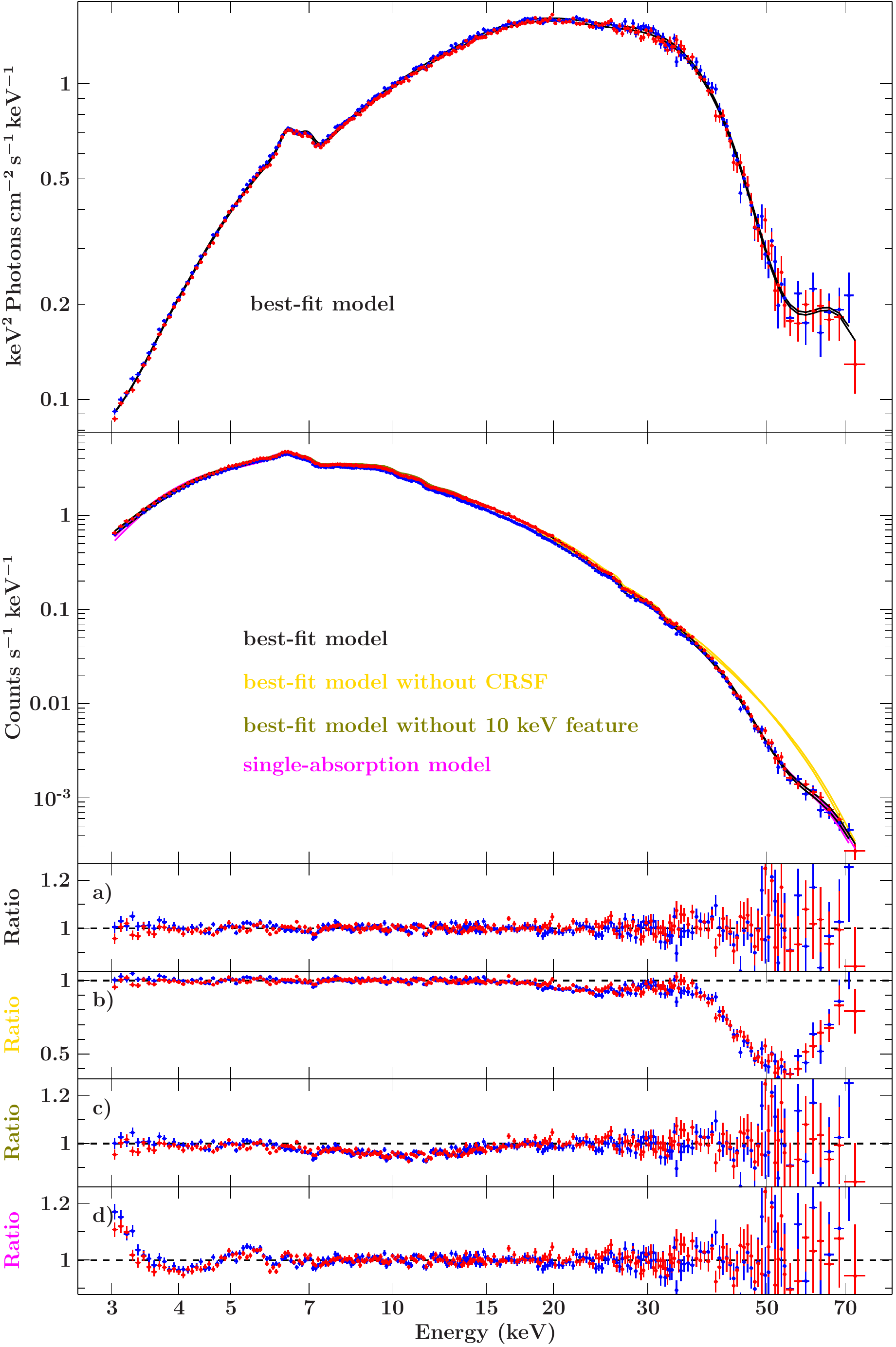}}
    \caption{Time-averaged spectrum for observation I. The FPMA data are in red and FPMB data in blue, and models for both focal plane modules are shown in the same colour and are hardly distinguishable with the naked eye. The first two panels show, respectively, the unfolded spectrum and the count spectrum. The residual panels are (from top to bottom): a) best-fit model (black), b) best-fit model with CRSFs (yellow) turned off but not fitted again, c) best-fit model with 10\,keV feature (green) turned off but not fitted again, and d) single-absorption model (magenta). Green and magenta lines are almost hidden by the data and the best-fit model in the data panel.}
    \label{fig:step-by-step_Felix}
\end{figure}

\begin{figure*}[!htpb]
\begin{subfigure}{.5\textwidth}
    \centering
    \includegraphics[trim=0cm 0cm 0cm 0cm, clip=true, width=1.0\linewidth]{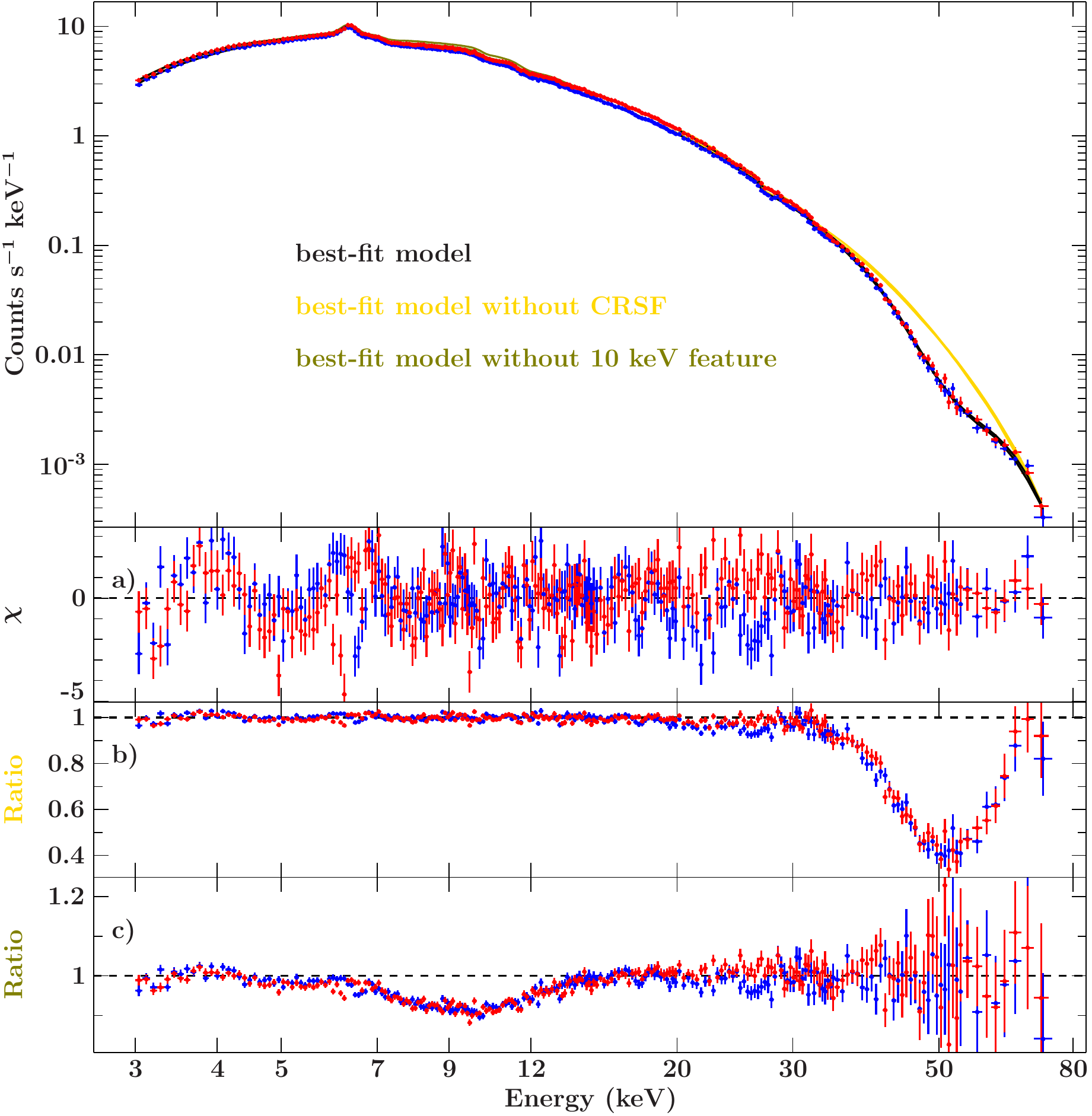}
    \label{fig:step-by-step_Vic_before}
    \end{subfigure} 
    \begin{subfigure}{.5\textwidth}
    \centering
    \includegraphics[trim=0cm 0cm 0cm 0cm, clip=true, width=1.0\linewidth]{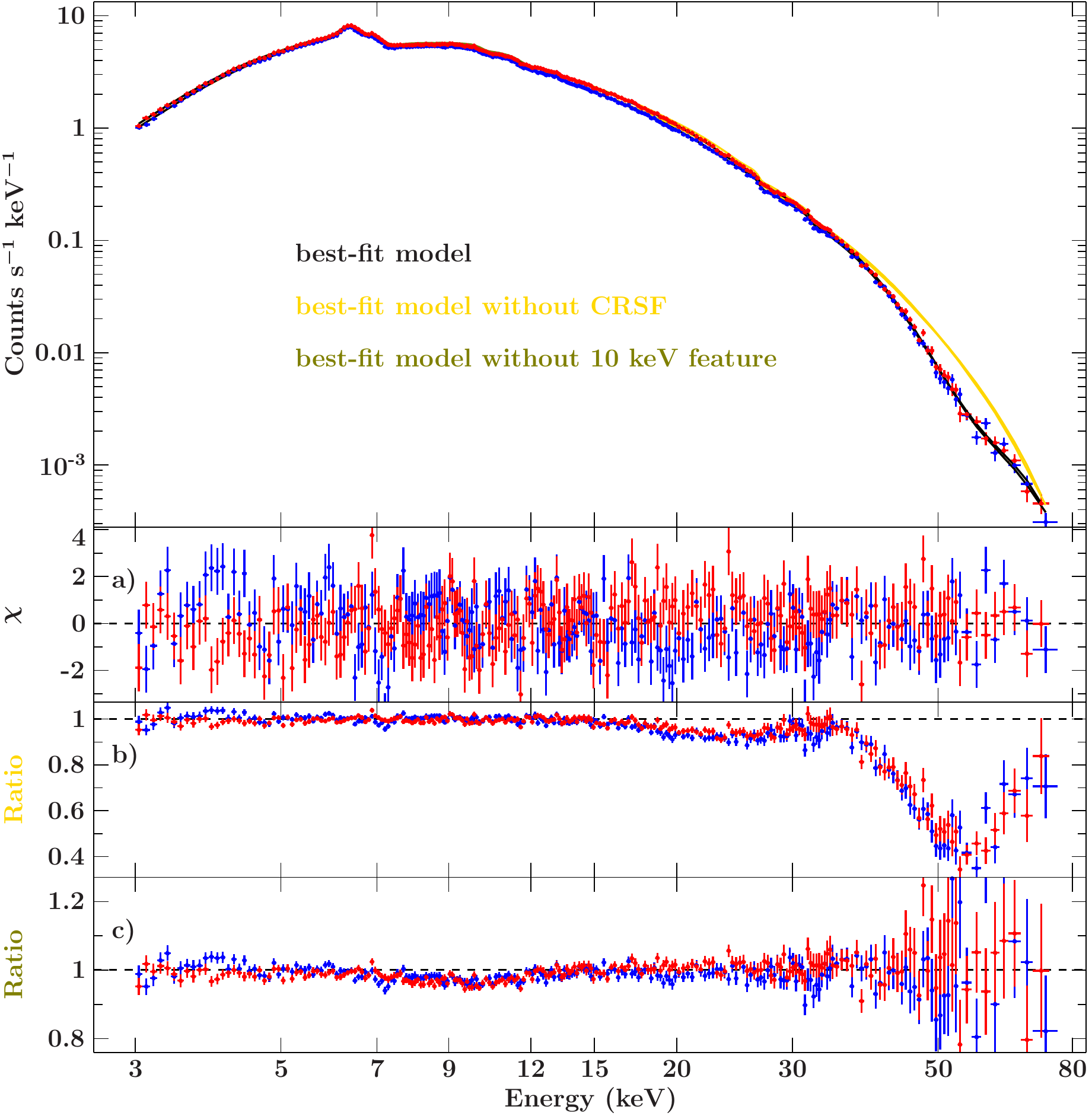}
    \label{fig:step-by-step_Vic_after}
    \end{subfigure}
    \caption{Time-averaged spectrum for observations IIa (\textsl{left panel}) and IIb (\textsl{right panel}). The FPMA data are in red and the FPMB data in blue.\ Models for both focal plane modules are shown in the same colour and are hardly distinguishable with the naked eye. The residual panels are (from top to bottom): a) best-fit model (black), b) best-fit model with CRSFs (yellow) turned off but not fitted again, c) best-fit model with 10\,keV feature (green; this line is almost hidden by the data and the best-fit model in the data panel) turned off but not fitted again.}
    \label{fig:step-by-step_Vic}
\end{figure*}

The best-fit parameters for the three observations are presented in Table \ref{tab:average_bestfit}. There is an increase in the stellar wind absorption $N_{\rm{H,1}}$ between observations IIa and IIb together with an increase in the covering fraction $\rm{CF}$. Even though the energy of the fundamental CRSF $E_{\rm{CRSF,F}}$ remains almost constant, we can notice variability in the energy of the harmonic CRSF $E_{\rm{CRSF,H}}$ between the three observations. The folding energy $E_{\rm{fold}}$ remains constant within the uncertainties between the three observations whereas we can observe a variable cutoff energy $E_{\rm{cut}}$. The unabsorbed flux $\mathcal{F}_{\rm{3-79 \ keV}}$ is higher by a factor of 2 in observations IIa and IIb than in observation I, following the higher count rates observed. The comparatively poor fit in observation IIa is driven by the low energies, where the spectrum appears even more complex than the model assumes (Fig.~\ref{fig:step-by-step_Vic}).

\begin{table*}[!htpb]
\renewcommand{\arraystretch}{1.1}
\caption{Best-fit parameters for the time-averaged final model.}
\label{tab:average_bestfit}
\begin{center}
\begin{small}
\begin{tabular}{llllll}    
\hline\hline
Parameter & 
Obs I &
Obs IIa  &
Obs IIb &
Orbit-by-orbit &
Pulse-by-pulse \\
\hline
$C_{\rm{FPMA}}$  &  1  &  1  &  1  &  1  &  1\\
$C_{\rm{FPMB}}$  &  $1.0203\pm0.0021$  & $1.0142\pm0.0018$ & $1.0204\pm0.0022$  &  $\Rightarrow$ fixed  &  $\Rightarrow$ fixed\\
$N_{\rm{H,1}} \ (10^{22} \,  \rm{cm^{-2}})$  &  $34.0\pm1.0$  &  $32.1\pm1.9$ & $42.8^{+1.2}_{-1.3}$  &  fitted  &  fitted\\
$N_{\rm{H,2}} \ (10^{22} \, \rm{cm^{-2}})$  &  fixed to 0.371  &  fixed to 0.371 & fixed to 0.371 & fixed to 0.371 & fixed to 0.371\\
$\Gamma$  &  $1.09\pm0.05$  & $1.084^{+0.024}_{-0.025}$ & $1.01^{+0.04}_{-0.05}$ & fitted & fitted \\
$E_{\rm{cut}} \ (\rm{keV})$  &  $19.9^{+3.5}_{-2.0}$  &  $26.1^{+1.1}_{-1.2}$ & $21\pm4$ & $\Rightarrow$ fixed & $\Rightarrow$ fixed\\
$E_{\rm{fold}} \ (\rm{keV})$  &  $12.6^{+0.9}_{-0.8}$  &  $10.2^{+0.5}_{-0.4}$ & $11.2^{+1.1}_{-2.1}$ & fitted & fitted\\
$E_{\rm{CRSF,F}} \ (\rm{keV})$  &  $24.7^{+1.0}_{-0.9}$  & $24.3^{+1.0}_{-1.1}$ & $24.0^{+0.0}_{-0.8}$ & fitted & $\Rightarrow$ fixed\\
$\sigma_{\rm{CRSF,F}} \ (\rm{keV})$  &  fixed to $0.5\times \sigma_{\rm{CRSF,H}}$  & fixed to $0.5\times \sigma_{\rm{CRSF,H}}$ & fixed to $0.5\times \sigma_{\rm{CRSF,H}}$ & $\Rightarrow$ fixed  & $\Rightarrow$ fixed\\
$d_{\rm{CRSF,F}} \ (\rm{keV})$  &  $0.75^{+0.00}_{-0.19}$  & $0.31^{+0.16}_{-0.11}$ & $0.86^{+0.00}_{-0.24}$ & fitted & $\Rightarrow$ fixed\\
$E_{\rm{CRSF,H}} \ (\rm{keV})$  &  $53.8^{+1.1}_{-0.9}$  & $51.8\pm0.7$ & $56.0^{+2.0}_{-1.2}$ & fitted & $\Rightarrow$ fixed\\
$\sigma_{\rm{CRSF,H}} \ (\rm{keV})$  &  $7.9^{+1.3}_{-0.9}$  & $7.4^{+0.8}_{-0.6}$ & $8.8^{+6.6}_{-1.2}$ & $\Rightarrow$ fixed & $\Rightarrow$ fixed\\
$d_{\rm{CRSF,H}} \ (\rm{keV})$  &  $18^{+6}_{-4}$  & $16.5^{+3.1}_{-2.5}$ & $18.58^{+0.00}_{-4.65}$ & fitted & $\Rightarrow$ fixed\\
$E_{\rm{FeK\alpha}} \ (\rm{keV})$  &  $6.364\pm0.012$  & $6.357\pm0.006$ & $6.356\pm0.008$ & fitted & fitted\\
$A_{\rm{FeK\alpha}} \ (\rm{ph \, s^{-1} \, cm^{-2}})$  &  $\left(1.35\pm0.11\right)\times10^{-3}$ & $\left(4.04^{+0.17}_{-0.16}\right)\times10^{-3}$ & $\left(4.50^{+0.23}_{-0.22}\right)\times10^{-3}$ & fitted & fitted\\
$\sigma_{\rm{FeK\alpha}} \ (\rm{keV})$  &  $0.070^{+0.029}_{-0.045}$  & $0.045^{+0.023}_{-0.045}$ & $0.074^{+0.021}_{-0.025}$ & $\Rightarrow$ fixed & $\Rightarrow$ fixed\\
$E_{10\,\rm{keV}} \ (\rm{keV})$  &  $9.5^{+0.6}_{-1.0}$  & $9.13^{+0.14}_{-0.17}$ & $9.2^{+0.5}_{-0.6}$ & $\Rightarrow$ fixed & $\Rightarrow$ fixed\\
$A_{10\,\rm{keV}} \ (\rm{ph \, s^{-1} \, cm^{-2}})$  &  $\left(-4.8^{+2.2}_{-4.8}\right)\times10^{-3}$  & $-0.0125^{+0.0017}_{-0.0023}$ & $\left(-4.5^{+1.6}_{-2.4}\right)\times10^{-3}$ & fitted & fitted\\
$\sigma_{10\,\rm{keV}} \ (\rm{keV})$  &  $3.2^{+1.0}_{-0.7}$  & $2.34^{+0.22}_{-0.18}$ & $2.1^{+0.6}_{-0.5}$ & $\Rightarrow$ fixed & $\Rightarrow$ fixed\\
$\mathcal{F}_{3-79 \ \rm{keV}} \ (\rm{keV \, s^{-1} \, cm^{-2}})$  &  $3.43^{+0.10}_{-0.05}$  & $6.35^{+0.09}_{-0.00}$ & $6.24^{+2.89}_{-0.10}$ & fitted & fitted \\
$\mathcal{F}_{3-79 \ \rm{keV}} \ (\rm{erg \, s^{-1} \, cm^{-2}})$  &  $\left(5.49^{+0.17}_{-0.08}\right) \times 10^{-9}$  & $\left(10.2^{+0.1}_{-0.00}\right) \times 10^{-9}$ & $\left(10.0^{+4.4}_{-0.4}\right) \times 10^{-9}$ & fitted & fitted \\
$\rm{CF}$  &  $0.883\pm0.005$  & $0.505^{+0.012}_{-0.013}$ & $0.852\pm0.005$ & fitted & fitted\\
$\chi^2/\rm{dof}$  &  $613.37/456$  & $731.94/456$ & $578.30/456$ \\
\hline
\multicolumn{6}{p{0.9\linewidth}}{In the two right-most columns, we indicate which spectral parameters are fixed to the value found in the corresponding analysis at lower time resolution or still left free during the fit.}
\end{tabular}
\end{small}
\end{center}
\end{table*}

\section{Time-resolved spectroscopy}
\label{section:timeres_spectro}

Now that we have a good description of the average continuum of Vela X-1 for our observations, we can use the model to access the variability of the source on shorter timescales. The good sensitivity of \nustar enables us to extract a spectrum for each \nustar orbit and also for each rotation of the neutron star. However, an analysis on a shorter timescale than the pulse period is not expedient because the spectrum is highly variable with the pulse phase \citep[see][]{Kreykenbohm_2002, LaBarbera_2003, Maitra&Paul_2013}.

We used the model in Eq.~\ref{eq:final_model} for the time-resolved spectroscopy, employing the same parameter ranges that we used for the time-averaged spectroscopy. However, given the shorter exposure time per spectrum and consequently lower signal-to-noise ratio, we have to fix some more parameters (see Table~\ref{tab:average_bestfit}) to their time-averaged values to obtain meaningful constraints on our model parameters. As the background is stable during the observation, we use the time-averaged background for all of the sliced-data since it provides a better estimate of the true background spectrum due to the higher signal-to-noise. 

Due to its proximity to and thus degeneracy with $E_{\rm{CRSF,F}}$, the $E_{\rm{cut}}$ parameter has to be fixed for the orbit-by-orbit analysis. We further fix the line widths of both harmonic and fundamental CRSFs, $\sigma_{\rm{CRSF,H}}$ and $\sigma_{\rm{CRSF,F}}$, the line width of the $\rm{FeK\alpha}$ line, $\sigma_{\rm{FeK\alpha}}$, the energy of the 10\,keV feature, $E_{10\,\rm{keV}}$, the line width of the 10\,keV feature, $\sigma_{10\,\rm{keV}}$, and the floating cross-normalisation parameter, $C_{\rm{FPMB}}$, to their respective time-averaged values.

\begin{figure*}
    \centerline{\includegraphics[trim=0cm 0cm 0cm 0cm, clip=true, width=0.8\linewidth]{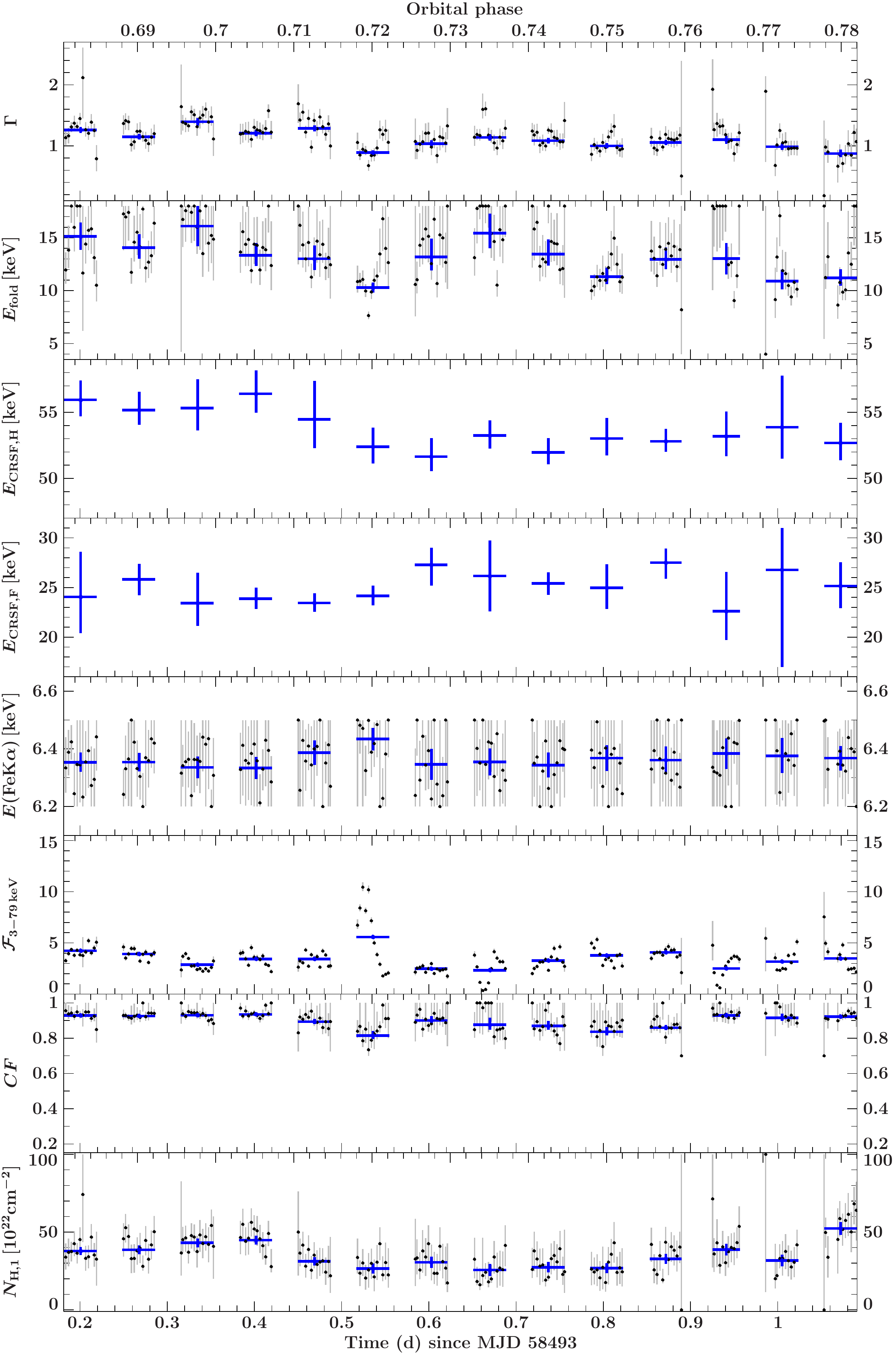}}
    \caption{Results of the pulse-by-pulse (black) and orbit-by-orbit (blue) analyses for observation I as functions of time, showing also the corresponding orbital phase. The panels show (from top to bottom): photon index ($\Gamma$), folding energy ($E_{\rm{fold}}$) in keV, energy of the harmonic CRSF ($E_{\rm{CRSF,H}}$) in keV, energy of the fundamental CRSF ($E_{\rm{CRSF,F}}$) in keV, energy of the $\rm{FeK\alpha}$ line in keV, unabsorbed flux ($\mathcal{F}_{\rm{3-79 \ keV}}$) in $\rm{keV \, s^{-1} \, cm^{-2}}$, covering fraction ($\rm{CF}$), and absorption from the stellar wind ($N_{\rm{H,1}}$) in $10^{22}\,\rm{cm^{-2}}$.}
    \label{fig:orb-pulse_time_Felix}
\end{figure*}

\begin{figure*}
    \centerline{\includegraphics[trim=0cm 0cm 0cm 0cm, clip=true, width=0.8\linewidth]{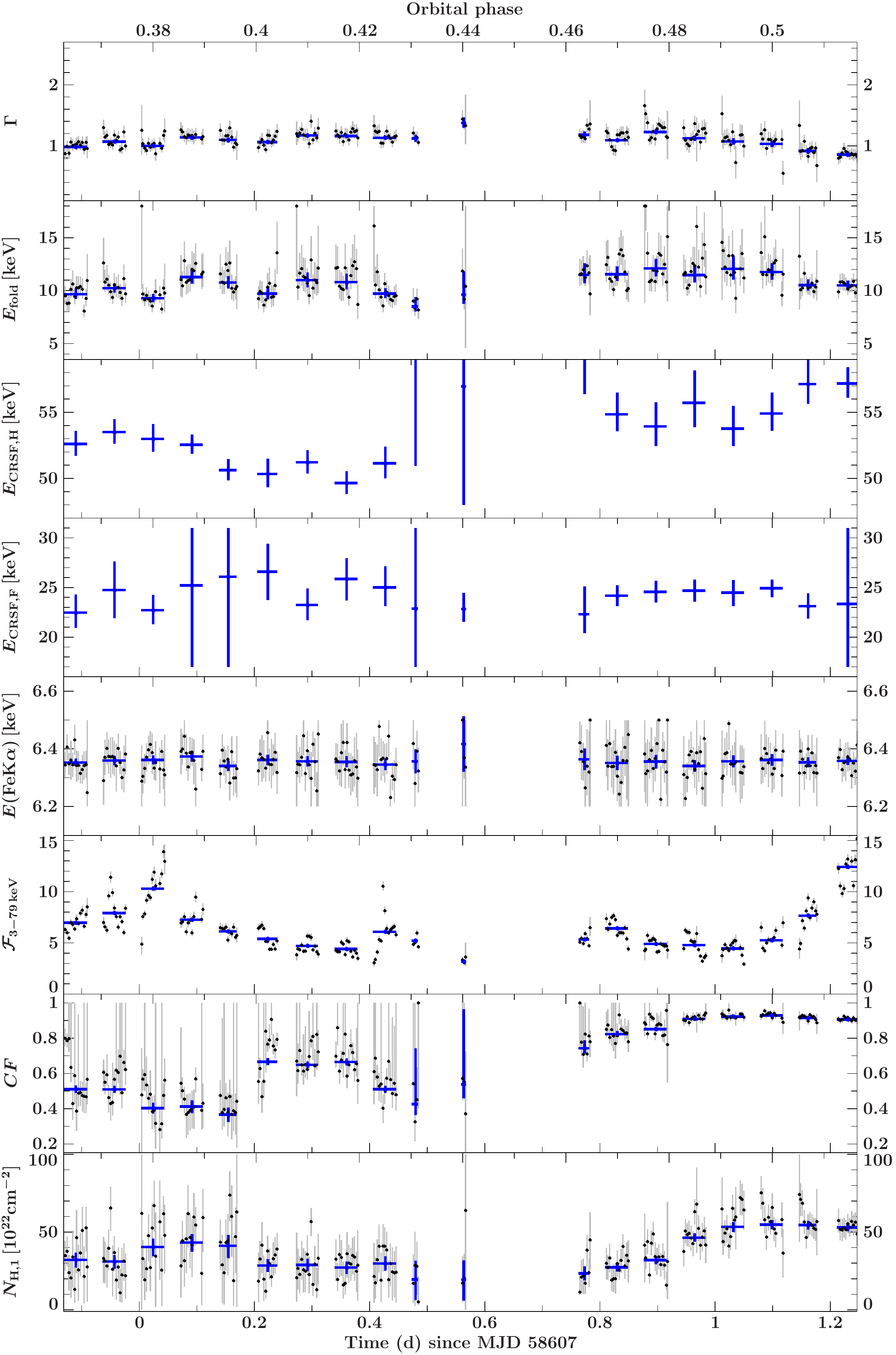}}
    \caption{Results of the pulse-by-pulse (black) and orbit-by-orbit (blue) analyses for observation II as functions of time, showing also the corresponding orbital phase. The panels show (from top to bottom): photon index ($\Gamma$), folding energy ($E_{\rm{fold}}$) in keV, energy of the harmonic CRSF ($E_{\rm{CRSF,H}}$) in keV, energy of the fundamental CRSF ($E_{\rm{CRSF,F}}$) in keV, energy of the $\rm{FeK\alpha}$ line in keV, unabsorbed flux ($\mathcal{F}_{\rm{3-79 \ keV}}$) in $\rm{keV \, s^{-1} \, cm^{-2}}$, covering fraction ($\rm{CF}$), and absorption from the stellar wind ($N_{\rm{H,1}}$) in $10^{22}\,\rm{cm^{-2}}$.}
    \label{fig:orb-pulse_time_Victoria}
\end{figure*}

For the pulse-by-pulse spectroscopy, we start with the same settings and fixed parameters as for the orbit-by-orbit spectroscopy. However, several of the remaining parameters are not well constrained by these data. Thus, we have to fix the strengths $d_{\rm{CRSF,H}}$ and $d_{\rm{CRSF,F}}$ and energies $E_{\rm{CRSF,H}}$ and $E_{\rm{CRSF,F}}$ of both harmonic and fundamental CRSFs to their respective orbit-wise values. We show the results of both orbit-by-orbit and pulse-by-pulse spectral analyses as functions of time and orbital phase for observations I and II (IIa--IIb) in Figs.~\ref{fig:orb-pulse_time_Felix} and~\ref{fig:orb-pulse_time_Victoria}, respectively.

Some of the individual values presented correspond to observations with very short exposures and can be usually recognised as outliers with large uncertainties. This is in particular the case for orbit-wise spectra at orbits 10, 11, and 12. These data have been affected by the loss of data due to ground station problems. Several pulse-wise outliers are located at the edges of the individual orbits where the data covers only a part of a given pulse. When discussing possible parameter correlations, these measurements have to be treated with caution. The off-states can be clearly seen in our data, but are too short for a detailed spectral analysis.

A significant variability of the presented parameters with time and orbital phase is observed in observations I and II. In particular, during the flare in observation I (at $T_{\rm{obs}} \approx 58493.53$\,MJD), $\Gamma$ reaches its minimal value while the energy of the FeK$\alpha$, which is rather stable otherwise, increases.

Significant variations in the energy of CRSFs and especially $E_{\rm{CRSF,H}}$ can be observed in both datasets for the orbit-by-orbit spectral analysis. In observation IIb the values of $E_{\rm{CRSF,H}}$ are higher than during observation IIa. In observation I, $E_{\rm{CRSF,H}}$ is at higher energies before than after the flare. The fundamental line is much weaker than the harmonic and thus the uncertainties on the values are much larger, preventing us from drawing conclusions on any correlations.

The stellar wind absorption is high in both observations. $N_{\rm{H,1}}$ in observation I averages around $\sim$40$\times10^{22}\,\rm{cm^{-2}}$. In observation II, it is more variable, in particular during observation IIb: it shows a rapid increase to more than $\sim$50$\times10^{22}\,\rm{cm^{-2}}$.

The covering fraction $\rm{CF}$ remains stable around 0.9 for observation I. In observation II, $\rm{CF}$ is more variable, ranging between $\sim$0.3 and $\sim$1 and is higher during observation IIb than during observation IIa. During flares, the covering fraction tends to decrease on the pulse-by-pulse timescale as visible in observation I at $T_{\rm{obs}} \approx 58493.53$\,MJD or in observation II at $T_{\rm{obs}} \approx 58606.95$\,MJD, at $T_{\rm{obs}} \approx 58607.03$\,MJD and at $T_{\rm{obs}} \approx 58607.42$\,MJD. During off-states ($T_{\rm{obs}} \approx 58493.66$\,MJD and at $T_{\rm{obs}} \approx 58493.93$\,MJD in observation I), which corresponds to a minimum in $\mathcal{F}_{3-79\,keV}$, the $\rm{CF}$ tends to reach its maximal value of 1. 

The folding energy $E_{\rm{fold}}$ seems to show distinct behaviour during both the flare and the off-states, but a correlation cannot be easily concluded without further investigation (see Sect.~\ref{section:discussion}).

\section{Discussion}
\label{section:discussion}

\subsection{Spectral shape and CRSF variability with luminosity}

In this subsection, to compare our results to previous work and to other sources, we have to convert our observed 3--79\,keV flux, $\mathcal{F}_{\rm{3-79\,keV}}$, to the 3--79\,keV luminosity, $L_{\rm{3-79\,keV}}$, using $\mathcal{F}_{\rm{3-79\,keV}} = {L_{\rm{3-79\,keV}}}/{(4\pi d^2),}$
where $d$ is the distance to the source and is equal to $1.99^{+0.13}_{-0.11}$\,kpc \citep{Kretschmar_2021a}. Overall, our luminosities are in the range of \mbox{$\sim$2--$10 \times 10^{36}\,\mathrm{erg}\,\mathrm{s}^{-1}$} (Fig.~\ref{fig:furst2014_fig7}) and thus slightly but not drastically below the often quoted range of around or a few $10^{37}\,\mathrm{erg}\,\mathrm{s}^{-1}$ for the critical luminosity \citep[e.g.][]{Reig_2013}.
Uncertainties on the distance lead to systematic uncertainties in the overall luminosity of the source of approximately +15\%/-12\%. For this work, we chose to use an empirical model to describe the spectrum of Vela X-1 in order to get insights into the underlying physics, which is discussed below.

\subsubsection{Continuum spectral shape}
\label{section:continuum_spectral_shape}

\begin{figure}[!tb]
    \centering
    \centerline{\includegraphics[trim=0cm 0cm 0cm 0cm, clip=true, width=1\linewidth]{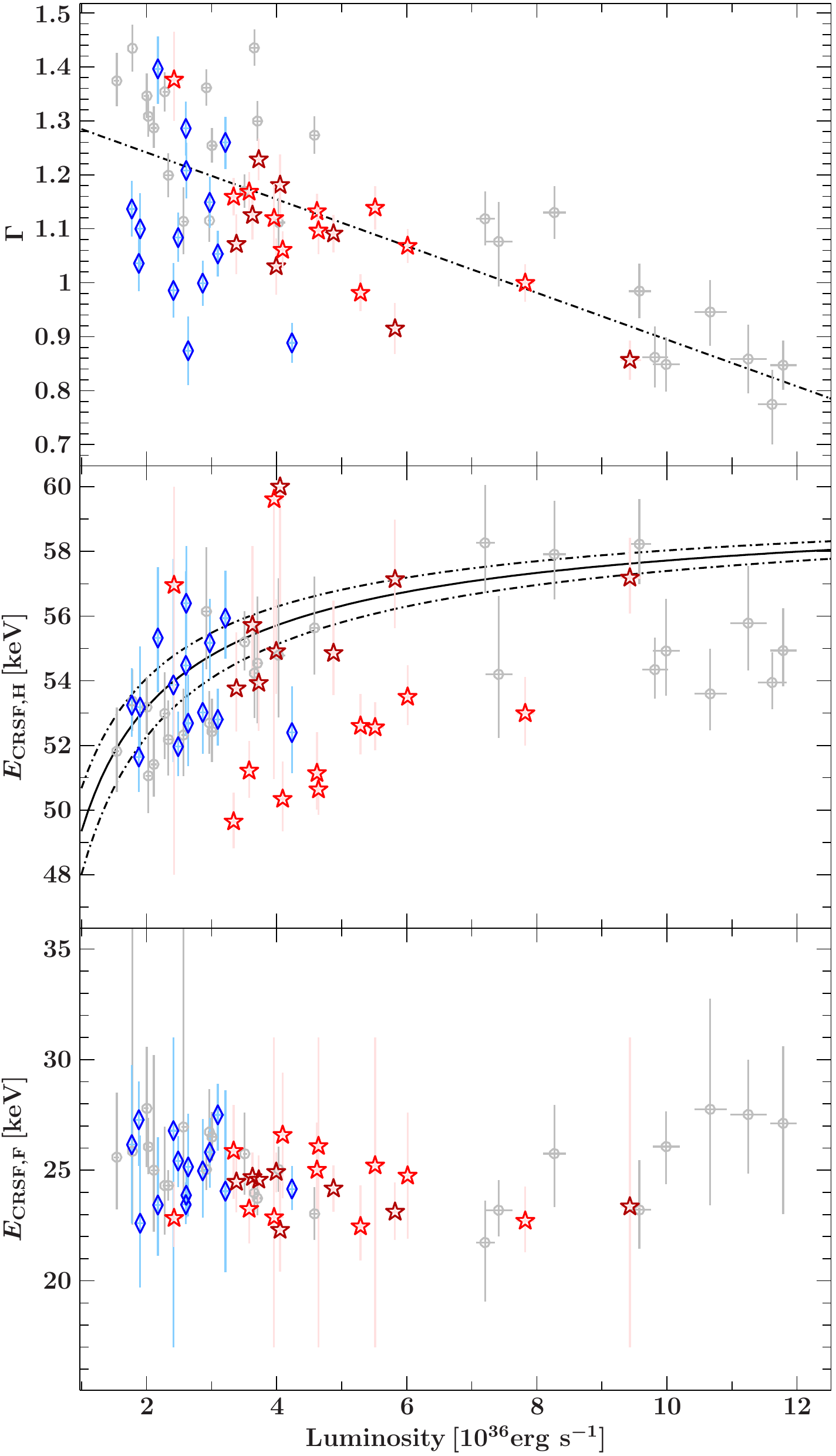}}
    \caption{Spectral parameter as a function of the 3--79\,keV luminosity based on our orbit-by-orbit spectroscopy results (dark blue diamonds represent observation I, bright red stars  observation IIa, and dark red stars observation IIb) together with kilosecond-integrated spectral fits from Fig.~7 in \citet{Fuerst_2014a} (grey circles). \textsl{Upper panel}: Photon index ($\Gamma$). The dot-dashed black line is a linear regression through all data points, meant to guide the eye. \textsl{Middle panel}: Energy of the harmonic CRSF ($E_{\rm{CRSF,H}}$). The solid black line is the theoretical prediction for $\Lambda = 1$, $E_{\rm{NS}}=30\,\rm{keV}$, and $M_{\rm{NS}} = 1.9\,M_{\odot}$. The dot-dashed lines above and below  the solid line are for $M_{\rm{NS}} = 1.7\,M_{\odot}$ and $M_{\rm{NS}} = 2.1\,M_{\odot}$, respectively. \textsl{Lower panel}: Energy of the fundamental line ($E_{\rm{CRSF,F}}$).}
    \label{fig:furst2014_fig7}
\end{figure}

In the upper panel of Fig.~\ref{fig:furst2014_fig7}, we show our results for the correlation between $\Gamma$ and the luminosity. We also include previous measurements by \citet{Fuerst_2014a} at orbital phase 0.655--0.773, re-scaled to the updated distance to the source.
We can observe a negative correlation between the spectral slope and the luminosity. This behaviour is typical for a source in the subcritical accretion regime as it has already been found for example for Her X-1 \citep{Staubert_2007}, Vela X-1 \citep{Odaka_2013,Fuerst_2014a}, and several Be/X-ray pulsars in the subcritical accretion regime \citep{Reig_2013}. Assuming X-ray radiation dominated by Comptonisation, this correlation can be a consequence of the increase in the accretion rate resulting in an increase in the X-ray luminosity and to a more efficient Comptonisation occurring in the accreted plasma \citep{Odaka_2013}. 

\begin{figure}[!tb]
    \centering
    \centerline{\includegraphics[trim=0cm 0cm 0cm 0cm, clip=true, width=1.\linewidth]{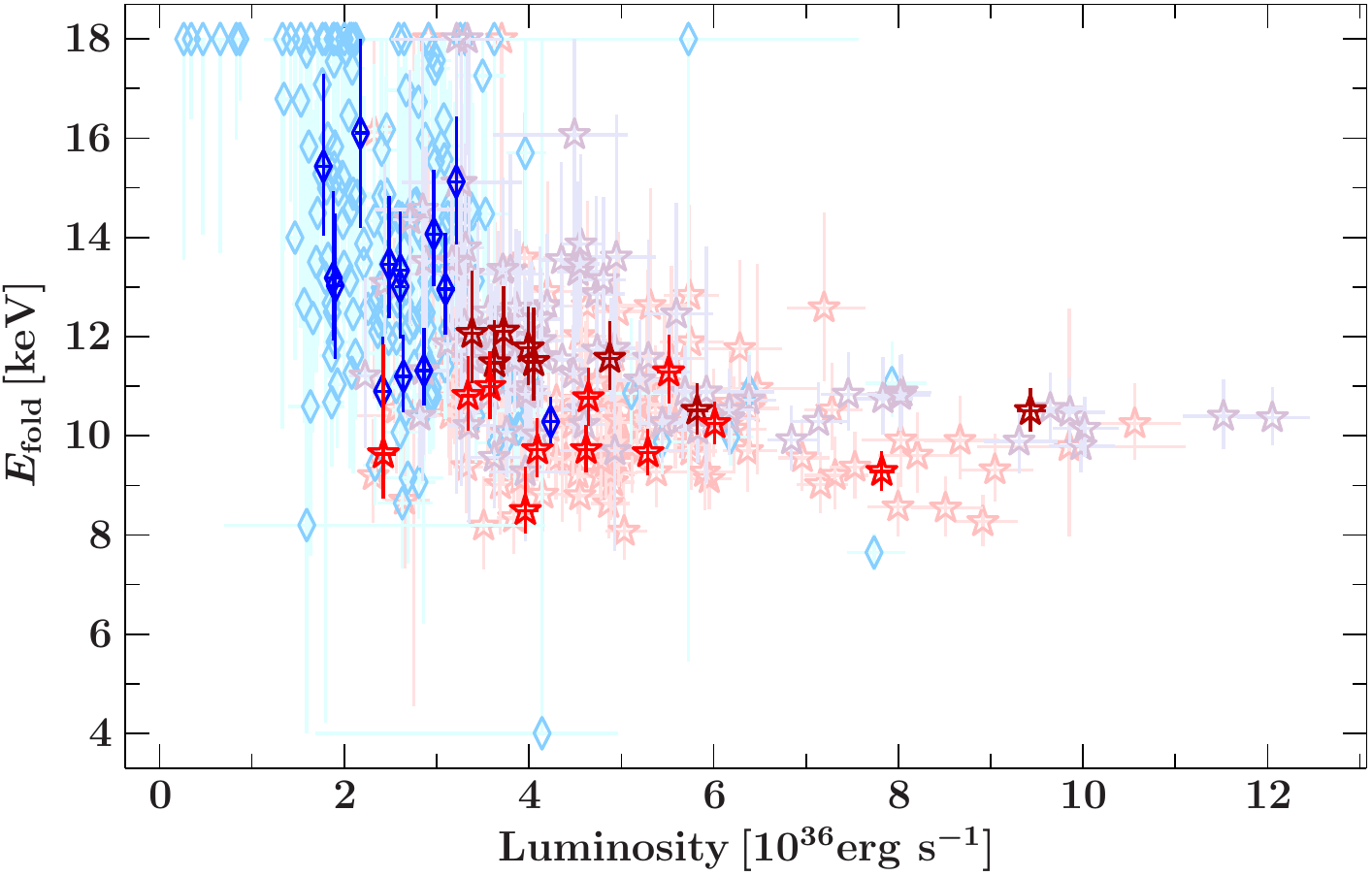}}
    \caption{Folding energy ($E_{\rm{fold}}$) as a function of the 3--79\,keV luminosity based on our orbit-by-orbit and pulse-by-pulse spectroscopy results. Dark blue diamonds are the results of the orbit-by-orbit analysis for observation I, light blue diamonds are the results of the pulse-by-pulse spectroscopy for observation I, bright red stars are the results of the orbit-by-orbit spectroscopy for observation IIa, pink stars are the results of the pulse-by-pulse spectroscopy for observation IIa, dark red stars are the results of the orbit-by-orbit spectroscopy for observation IIb, and lavender stars are the results of the pulse-by-pulse spectroscopy for observation IIb.}
    \label{fig:foldE_vs_lum}
\end{figure}

In Fig.~\ref{fig:foldE_vs_lum} we can also see an anti-correlation between $E_{\rm{fold}}$ and luminosity, as already indicated in Figs.~\ref{fig:orb-pulse_time_Felix} and \ref{fig:orb-pulse_time_Victoria} for lower luminosities ($\lesssim 4\times10^{36}\,\rm{erg\,s^{-1}}$). For higher luminosities, the folding energy seems to be constant around $\sim$10\,keV.  Considering the underlying shape of the model we employ, we expect a positive degeneracy between the folding energy $E_{\rm{fold}}$ and the X-ray flux (or luminosity). Confidence maps calculated for folding energy and flux indeed confirm this expectation, implying that the observed opposite trend cannot be explained by modelling degeneracies. \citet{Odaka_2013} use a different continuum model (\texttt{NPEX}) when analysing \textsl{Suzaku} data, so that the values of their and our parameters called folding energy cannot be directly compared. The trends, however, are comparable: they also find a negative correlation followed by a stabilisation between folding energy and flux at higher fluxes. Similar behaviour of the cutoff has been observed in a sample of Be/X-ray binaries when modelling their spectra with a cutoff power law in the subcritical accretion regime \citep{Reig_2013}. Assuming a significant contribution of not only bulk, but also thermal Comptonisation to the overall emission \citep{Becker_2007,Ferrigno_2009a}, the change in the cutoff would correspond to a change in the temperature of the Comptonising plasma.

\subsubsection{Accretion regime}
\label{section:accretion_regime}

To verify the accretion regime of Vela X-1 for our observations, we computed its critical luminosity using Eq. 32 from \citet{Becker_2012}:

\begin{align}
\label{eq:critical_lum}
\begin{split}
L_{\rm{crit}} = &1.49 \times 10^{37} {\rm{erg}}\,{\rm{s}}^{-1} \left(\frac{\Lambda}{0.1}\right)^{-7/5} \omega^{-28/15} \\
&\times \left(\frac{M_{\rm{NS}}}{1.4\,M_{\odot}}\right)^{29/30} \left(\frac{R_{\rm{NS}}}{10\,{\rm{km}}}\right)^{1/10}
\left(\frac{E_{{\rm{NS}},n}}{n\times11.57\,{\rm{keV}}}\right)^{16/15},
\end{split}
\end{align}
where $M_{\rm{NS}}$ and $R_{\rm{NS}}$ are the mass and radius of the neutron star and $\omega$ is a parameter describing the spectral shape inside the column. $E_{{\rm{NS}},n}$ is the surface cyclotron fundamental ($n=1$) or harmonic ($n=2$) energy of the neutron star. We use $\omega = 1$ (assuming that the spectrum inside the emission region is dominated by Bremsstrahlung). Recent publications \citep{Legred_2021} tend to assume a typical neutron star radius $R_{\rm{NS}}$ around 12\,km. However, for this work, we used $R_{\rm{NS}}=10\,\rm{km}$ under the same assumptions as in \citet{Becker_2012} for typical neutron star parameters \footnote{The  critical luminosity with a neutron star radius of 12\,km is $\sim$2\% higher than the critical luminosity using a neutron star radius of 10\,km.}. We considered a mass of $\sim$1.7--2.1$\,M_{\odot}$ \citep{Kretschmar_2021a} and a surface fundamental cyclotron energy of $E_{\rm{NS}}= 20\,\rm{keV}$. 

Parameter $\Lambda$ in Eq.~\ref{eq:critical_lum} accounts for the difference in location of the Alfvén surface in case of the predominantly orbital or radial inflow \citep[see, for example, Eq.~18 from][]{Lamb_1973}. It depends on the ratio of the radial velocity of the flow to the free-fall one and on the thickness of the accretion disc. However, the Alfvén radius is reduced only by a factor of $\sim$10 ($\Lambda \sim 0.1$) from the radius corresponding to a fully radial inflow even for the extreme values of these parameters (small radial velocities and a thin accretion disc). For spherical accretion, $\Lambda = 1$ and for disc accretion $\Lambda < 1$ \citep{Becker_2012}. In the case of Vela X-1, we consider a wind accreting source so the accretion is more likely to be spherical than through an accretion disc, although there are theoretical predictions of possible disc formation \citep{El_Mellah_2019a} and observational hints that temporarily a disc can be formed sometimes \citep{Liao_2020}. Additionally, even without the presence of a disc, the accretion from the clumpy, disturbed wind of the companion is more complex than a simple spherical accretion case \citep[e.g.][]{El_Mellah_2018a}.

If we consider a spherical accretion with $\Lambda = 1$, then $L_{\rm{crit}}\approx$~ 0.13--0.15$ \times 10^{37}\,\rm{erg}\,s^{-1}$ using Eq.~\ref{eq:critical_lum} and taking uncertainties on the mass into account.\ This classifies Vela X-1 as a super-critical source. 
However, in previous works, Vela X-1 has been classified as a subcritical source \citep{Fuerst_2014a} assuming $\Lambda = 0.1$ (thus disc accretion case), also following \citet{Becker_2012}. Indeed, for $\Lambda = 0.1$, we obtain $L_{\rm{crit}}\approx$~ 3.22--3.95$ \times 10^{37}\,\rm{erg}\,s^{-1}$. Our assumption of $E_{\rm{NS}} = 20$\,keV is conservative. Given that the measured energy of the fundamental line is above 20 keV, the real value is likely higher and will lead to a higher $L_{\rm{crit}}$. For instance, assuming $E_{\rm{NS}} = 30$\,keV, we obtain a $L_{\rm{crit}}\approx$~ 4.97--6.10$ \times 10^{37}\,\rm{erg}\,s^{-1}$. This approach puts Vela X-1 in the subcritical regime, which is supported by the observed anti-correlation between $\Gamma$ and luminosity (see Fig.~\ref{fig:furst2014_fig7}) typical for a subcritical source  \citep{Staubert_2007,Odaka_2013,Reig_2013,Fuerst_2014a}. However, the choice of $\Lambda = 0.1$ assumes a disc accretion case, which we question for our observations.

A further theoretical calculation for the critical luminosity has been presented in \citet{Mushtukov_2015a}, taking into account the exact Compton scattering cross-section in the high magnetic field and in particular they show the results of their calculations assuming $\Lambda = 0.5$. Their critical luminosity for Vela X-1 under these assumptions is of the order of $\sim$0.1--$1\times10^{37}\,\rm{erg\,s^{-1}}$, which is around our measurements (Fig.~\ref{fig:furst2014_fig7}). 

For the Be X-ray binary GRO J1008-57, \citet{Kuehnel_2017} have computed the theoretically expected critical luminosities at the transition between different accretion regimes after \citet{Becker_2012}, \citet{Mushtukov_2015a}, and \citet{Postnov_2015} (see their Appendix A). Uncertainties in the theories and data limitations do not allow them to favour one of the theories for the prediction of the critical luminosity, similar to our inconclusive results. 

Given the above and the fact that neither the disc accretion scenario nor the spherical accretion are a good description for the accretion in the highly structured, disturbed wind in the system, estimates on whether Vela X-1 is on sub- or supercritical regime have to be treated with caution. 
However, for the following section, Sect.~\ref{section:positive_correlation_CRSF_lum}, we assume a subcritical accretion regime for Vela X-1 to compare with previous results.

\subsubsection{CRSFs, luminosity, and flares}
\label{section:positive_correlation_CRSF_lum}

As we found a negative correlation between $\Gamma$ and luminosity for our observations in Sect.~\ref{section:continuum_spectral_shape}, we expect a positive correlation between the CRSF energy and the luminosity (see details in Sect.~\ref{section:intro}) in the case of a subcritical accretion regime.

The theoretical expected energy of the fundamental CRSF for a subcritical source can be calculated using Eq.~7 from \citet{Fuerst_2014a} with $\tau_*$, the Thomson optical depth, which we set to 20, the estimate by \citet{Becker_2012} for the plasma braking by Coulomb collisions in the filled accretion column and $E_{{\rm{NS}}}$ the energy of the fundamental cyclotron energy:
\begin{equation}
\label{eq:e_theo_sub}
\begin{split}
E_{\rm{theo}}&=\left[1+0.6\left(\frac{R_{\rm{NS}}}{10\,{\rm{keV}}}\right)^{-\frac{13}{14}}\left(\frac{\Lambda}{0.1}\right)^{-1}\left(\frac{\tau_*}{20}\right) \left(\frac{M_{\rm{NS}}}{1.4\,M_{\odot}}\right)^{\frac{19}{14}}\right.\\&\left.\times\,\left(\frac{E_{\rm{NS}}}{1\,{\rm{keV}}}\right)^{-\frac{4}{7}}\left(\frac{L_X}{10^{37}}\right)^{-\frac{5}{7}}\right]^{-3}\times E_{\rm{NS}}
\end{split}
.\end{equation}
To draw the prediction for the line energy of the harmonic CRSF ($n=2$) and to compare our results with \citet{Fuerst_2014a}, we have to use $2\times E_{\rm{theo}}$. The black dot-dashed line in Fig.~\ref{fig:furst2014_fig7} top panel shows the expected positive correlation between $E_{\rm{CRSF,H}}$ and luminosity in the case of a subcritical source using $\Lambda = 1$. However, this correlation is hardly visible for our data. This may be explained by the accretion regime of Vela X-1 lying somewhere in between the sub- and super-critical cases and thus yielding to inconclusive results. \citet{Vybornov_2017} also obtained inconclusive results for intermediate accretion regime case for Cep X-4 with collisionless shock theoretical study.

Overall, there seems to be an intrinsic variability in the CRSF energies in our observations but they do not appear to be correlated with the luminosity. In particular, in observation I, we observe a clear drop in CRSF energy from $\sim$55--56\,keV to $\sim$52--53\,keV after the flare even though the flux and other spectral parameters return to the pre-flare level (Fig.~\ref{fig:orb-pulse_time_Felix}). A similar decrease is seen in observation IIa (Fig.~\ref{fig:orb-pulse_time_Victoria}), but here the overall flux also decreases. If confirmed, such a behaviour could be explained by restructuring of the accretion column and thus change in the location of the CRSF producing region following an episode of stronger accretion.
To test whether the observed CRSF variability is real, we redid the orbit-by-orbit analysis, fixing the energies of the CRSFs to their respective average value. The overall fit gets worse compared to the analysis with those energies being free, and we have not seen the changes in the trend observed in other parameters.

\subsection{The variable absorption in the stellar wind}

\subsubsection{Orbital dependency: Wakes crossing our line of sight}

\begin{figure}
    \centering
    \centerline{\includegraphics[trim=6cm 7cm 0cm 14cm, clip=true, width=1.0\linewidth]{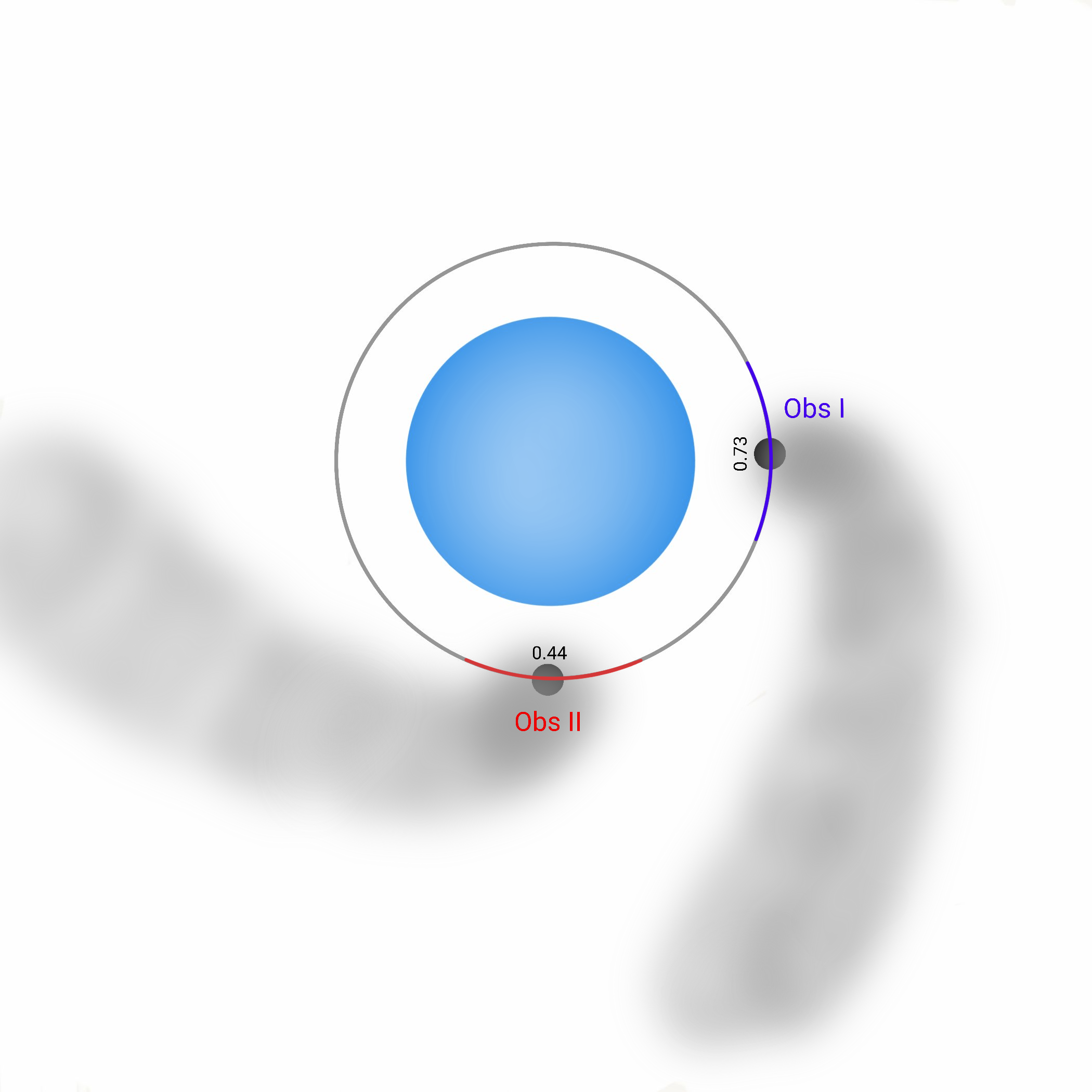}}
    \caption{Sketch of Vela X-1 showing possible positions of the wake structure during the two observations at mid-orbital phase. In this image, the observer's point of view is located facing the system at the bottom of the picture.}
    \label{fig:velax1_sketch}
\end{figure}

Because of the structure of the stellar wind, the amount of material being on our line of sight towards the neutron star can change and thus modifies the shape of the continuum creating variability. This is particularly striking with the covering fraction, $\rm{CF}$, and the equivalent absorbing hydrogen column density, $N_{\rm{H,1}}$, showing different behaviours between observations I and II (Figs.~\ref{fig:orb-pulse_time_Felix} and~\ref{fig:orb-pulse_time_Victoria}) that are taken at different orbital phases when our line of sight crosses different parts of the wind (Fig.~\ref{fig:velax1_sketch}).  The overall structure of the disturbed stellar wind is complex, with both photoionisation and accretion wake present and expected to be variable \citep{Blondin_1990a,Malacaria_2016a}; as our observations cannot distinguish between the components, we refer to them as wakes in general.

\begin{figure}
    \centering
    \centerline{\includegraphics[trim=0cm 0cm 0cm 0cm, clip=true, width=1\linewidth]{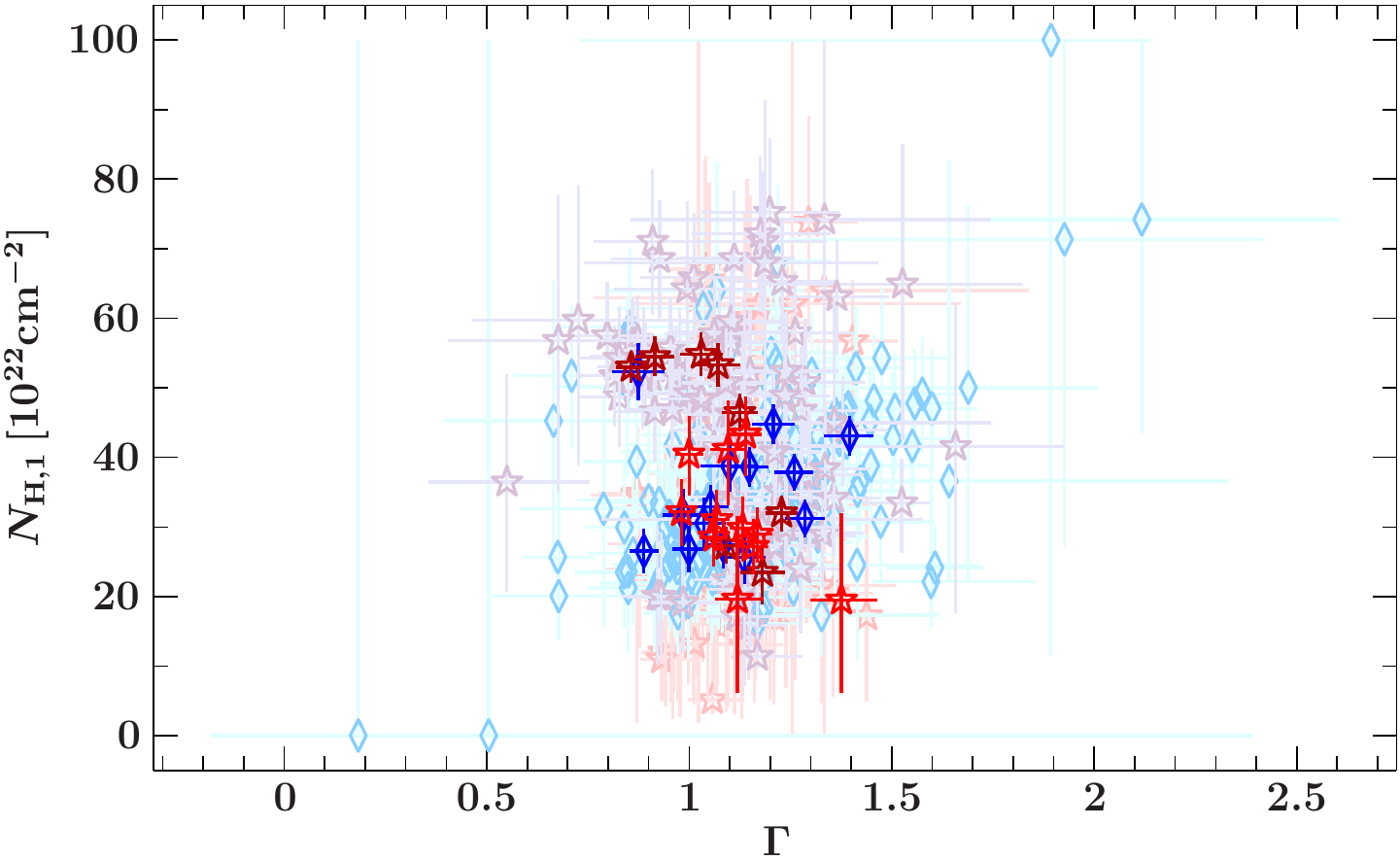}}
    \caption{Stellar wind absorption, $N_{\rm{H,1}}$, as a function of the photon index, $\Gamma$, based on our orbit-by-orbit and pulse-by-pulse spectroscopy results. Symbols are the same as in Fig.~\ref{fig:foldE_vs_lum}.}
    \label{fig:NH1_vs_Gamma}
\end{figure}

Thus, we explore possible correlations between the covering fraction and the column density. Additionally, there are possible modelling degeneracies such as between the slope, $\Gamma$, and $N_{\rm{H,1}}$, inherent to the usage of power law models for the continuum modelling of X-ray binaries \citep[e.g.][]{Suchy_2008a}. We carefully check for such a correlation both with scatter plots of our orbit-by-orbit and pulse-by-pulse fits (Fig.~\ref{fig:NH1_vs_Gamma}) and through calculation of confidence maps for $\Gamma$ and $N_{\rm{H,1}}$ and find that, even if present, it does not significantly contribute to our results.

\begin{figure}
    \centering
    \centerline{\includegraphics[trim=0cm 0cm 0cm 0cm, clip=true, width=1\linewidth]{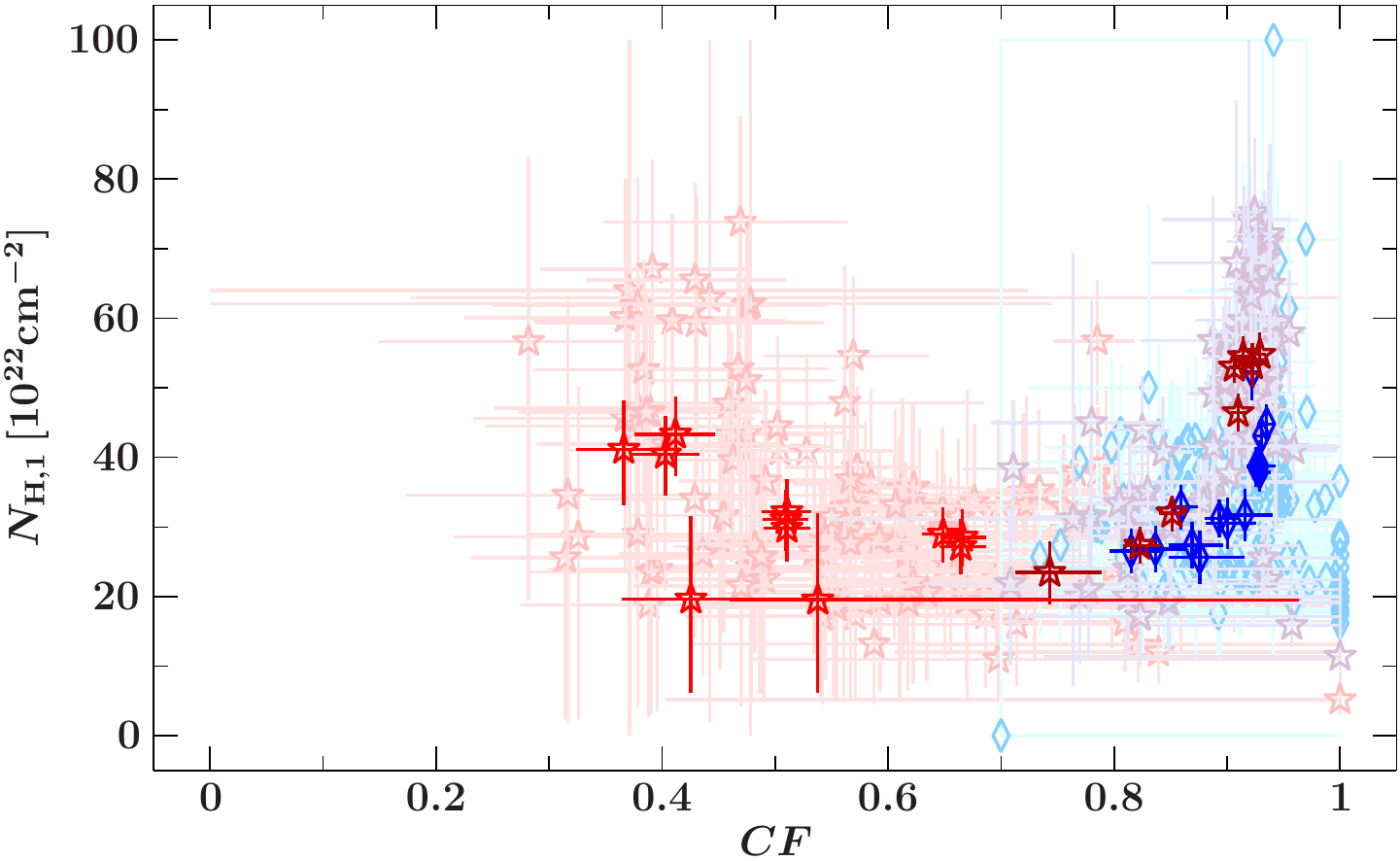}}
    \caption{Stellar wind absorption, $N_{\rm{H,1}}$, as a function of the covering fraction, $\rm{CF}$, based on our orbit-by-orbit and pulse-by-pulse spectroscopy results. Symbols are the same as in Fig.~\ref{fig:foldE_vs_lum}.}
    \label{fig:NH1_vs_CF}
\end{figure}

During observation I, the wakes are located on the line of sight towards the observer. The covering fraction is close to 1, meaning that most of the incoming photons are absorbed through the dense material in the absorption and photoionisation wakes \citep{Blondin_1990a,Kaper_1994a,Manousakis:PhD}. This is particularly visible in our time-resolved analysis in Fig.~\ref{fig:NH1_vs_CF} where the orbit-by-orbit and pulse-by-pulse data points of observation I show a positive correlation between the stellar wind absorption $N_{\rm{H,1}}$ and the covering fraction $\rm{CF}$. Similar high covering fractions and hydrogen column densities have been seen at this orbital phase in both \textsl{NuSTAR} \citep{Fuerst_2014a} and \textsl{Chandra} \citep{Amato_2021a} observations.

On the contrary, during observation IIa, the majority of the material in the wakes is not yet on the observer's line of sight, implying a low covering fraction $\rm{CF}$ and fewer photons absorbed by the stellar wind (Fig.~\ref{fig:velax1_sketch}). Indeed, the orbit-by-orbit and pulse-by-pulse data points of observation II in Fig.~\ref{fig:NH1_vs_CF} show a negative correlation between the stellar wind absorption $N_{\rm{H,1}}$ and the covering fraction $\rm{CF}$.

Finally, during observation IIb, the wakes are starting to pass through the observer's line of sight and consequently the $\rm{CF}$ slowly increases to reach the values of observation I (Fig.~\ref{fig:orb-pulse_time_Victoria}). This is confirmed by the time-resolved data points in Fig.~\ref{fig:NH1_vs_CF} where the same positive correlation than in observation I is seen.

To confirm the physical origin of these behaviours, we calculate the confidence maps for the two parameters of interest, $N_{\rm{H,1}}$ and $\rm{CF}$, using the results from the time-averaged spectroscopy for the three observations, I, IIa, and IIb. For observation I, pure modelling degeneracy would lead to a negative correlation between the two parameters, in opposite to what is seen on the scatter plot in Fig.~\ref{fig:NH1_vs_CF}, confirming the physical origin to the observed correlation.
Similarly, for observation IIa, modelling degeneracy would imply a positive correlation, while the time-resolved data points are showing a negative correlation between $N_{\rm{H,1}}$ and $\rm{CF}$ also confirming the real event. A possible physical explanation is the overlapping of multiple clumps. The stellar wind surrounding Vela X-1 is clumpy \citep[e.g.][]{Oskinova_2007} and most of the mass of the wind is concentrated in clumps \citep{Sako_1999a} so that some of them may be located on our line of sight during an observation. If one clump overlaps with another one, the absorption column density $N_{\rm{H,1}}$ increases and $\rm{CF}$ decreases as also observed by \citet{Fuerst_2014a}. For observation IIb, our approach does not detect an explicit modelling degeneracy between $N_{\rm{H,1}}$ and $\rm{CF}$.

\subsubsection{Absorption and luminosity}

\begin{figure}[!tb]
    \centering
    \centerline{\includegraphics[trim=0cm 0cm 0cm 0cm, clip=true, width=1.\linewidth]{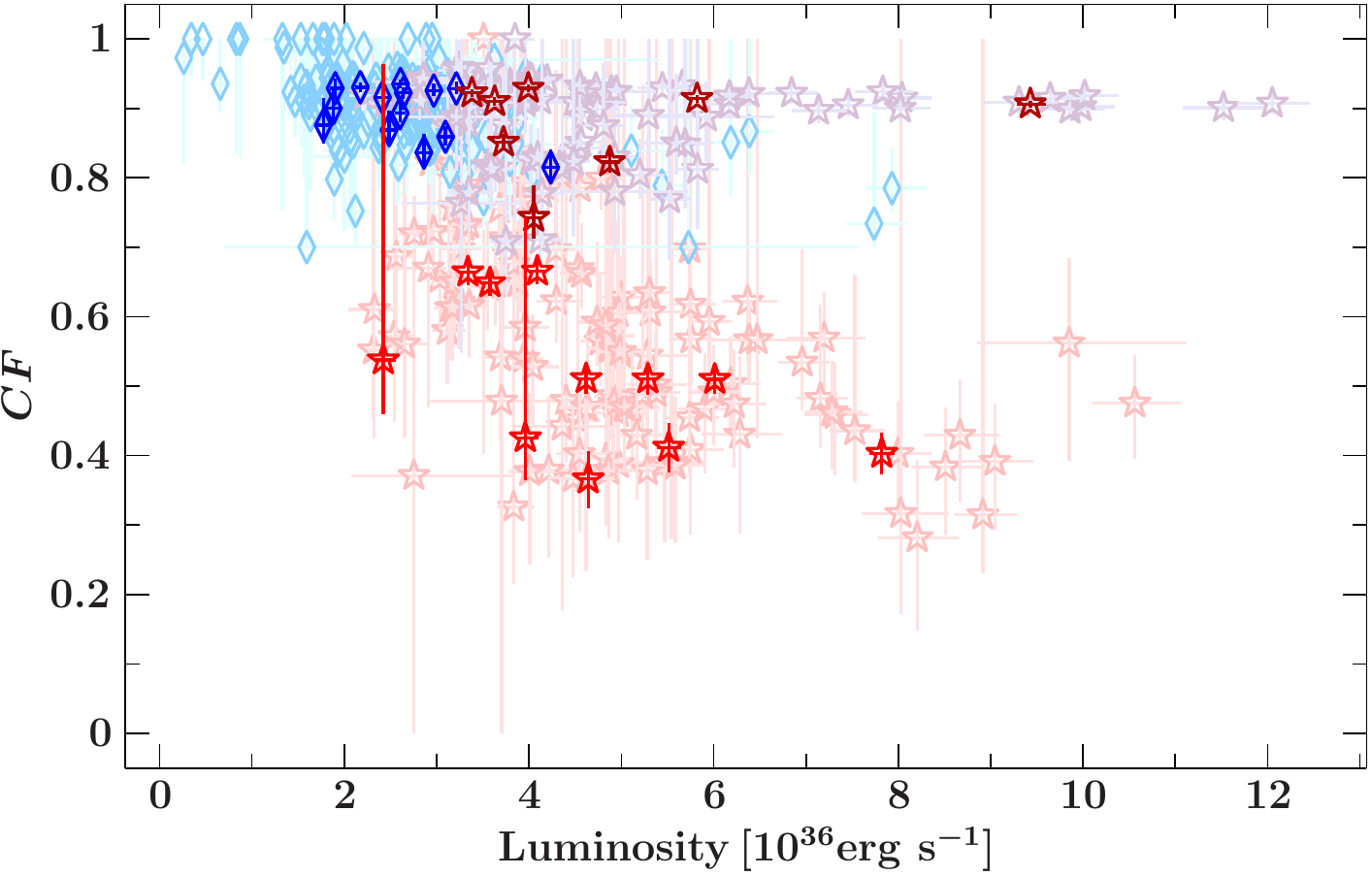}}
    \caption{Covering fraction, $\rm{CF}$, as a function of the 3--79\,keV luminosity based on our orbit-by-orbit and pulse-by-pulse spectroscopy results. Symbols are the same as in Fig.~\ref{fig:foldE_vs_lum}}
    \label{fig:CF_vs_lum}
\end{figure}

In observations I and IIb, when the wakes are located on the line of sight, Fig.~\ref{fig:CF_vs_lum} shows that the intrinsic luminosity of the source does not seem to be correlated with the covering fraction $\rm{CF}$. Because $\rm{CF}$ is correlated with $N_{\rm{H,1}}$ (Fig.~\ref{fig:NH1_vs_CF}) for these observations, we do not expect the luminosity of the source to correlate with $N_{\rm{H,1}}$ either. During a flaring or an off-state period, the fraction of photons absorbed through the stellar wind should be independent from the incoming number of photons if the dense and extended wakes are located on our line of sight as we do not expect them to be strongly influenced by the change in irradiation. However, there are some data points of observation I in Fig.~\ref{fig:CF_vs_lum} with high luminosities ($\sim$6--8$\times 10^{36}\,\rm{erg\,s^{-1}}$) and low $\rm{CF}$ ($\sim$0.7--0.8) that also correspond to a low $N_{\rm{H,1}}$. They correspond to a decrease in $\rm{CF}$ and $N_{\rm{H,1}}$ associated with flares in observations I and IIb (Fig~.~\ref{fig:orb-pulse_time_Felix} and \ref{fig:orb-pulse_time_Victoria}). This is in agreement with \citet{Odaka_2013}\footnote{Using the ephemeris listed in Table \ref{tab:ephemeris} and employed throughout this paper, the observation discussed in \citep{Odaka_2013} is at $\phi_{\rm{orb}}\approx$  0.20--0.38.} who have found with \textsl{Suzaku} observations that the circumstellar absorption does not seem to correlate with the X-ray luminosity except for a strong flare. 

\citet{Martinez_2014} have also observed a decrease in absorption during a giant flare that they attribute to the accretion of a dense clump in the wind. Our observations would support this scenario. Similar changes to the absorbing column density around and during flares have been seen in other wind-accreting sources, including in supergiant fast X-ray transients \citep{Pradhan_2019a,Bozzo_2017a}.

For observation IIa, the behaviour seems strikingly different (Fig.~\ref{fig:CF_vs_lum}), with possible indications of a negative correlation, although the large uncertainties preclude firm conclusions. Based on our knowledge of the geometry of the system, we interpret this as mainly a spurious correlation, linked to events that happen simultaneously in this particular observation by happenstance. At the beginning of observation IIa, the wakes are located outside our line of sight, which corresponds to a low covering fraction $\rm{CF}$. As seen in Fig.~\ref{fig:hr_lc}, the beginning of observation IIa also coincides with the first flaring period, where a deep local short decrease in $\rm{CF}$ (see Fig.~\ref{fig:orb-pulse_time_Victoria}) can also be observed for each flare due to the above described theory of accreted clumps. Similarly, while the wakes are starting to get towards our line of sight, the $\rm{CF}$ increases and this coincides with a decrease in the X-ray luminosity as the flare subsides (Fig.~\ref{fig:hr_lc}). 

The observed energy of the iron line is mostly constant throughout all discussed observations and consistent with neutral iron, except for one notable exception: during the bright flare in observation II, it increases (Fig.~\ref{fig:orb-pulse_time_Felix}). This could be indicative of increased ionisation during the flare. Interestingly, however, no increase in the iron line energy is seen in either parts of observation I (Fig.~\ref{fig:orb-pulse_time_Victoria}), nor was an increase in the iron line energy observed in a previous observation of a giant flare by \citet{Martinez_2014}. A careful modelling of our data during the flare does not reveal further possible correlations.

\section{Summary and outlook}
\label{section:summary_outlook}

We have analysed two new observations of Vela X-1 taken with \nustar at different orbital phases, thus probing a different line of sight through the system. For spectral modelling, we used a partial covering model with a power law continuum with a Fermi-Dirac cutoff further modified by CRSFs. We interpreted our observations in the context of a highly magnetised neutron star accreting from the clumpy, disturbed wind of a companion and draw conclusions about the accretion geometry and stellar wind structure and their interplay.

Our observations can be explained by denser, extended wakes passing through the line of sight starting at orbital phase $\sim$0.45. At lower orbital phases, the continuum is less absorbed by the stellar wind. The observed flares and the absorption variability during and around the flaring episodes can be explained by the accretion of clumps. It would be interesting, in future work, to perform a uniform analysis of Vela X-1 at all available orbital phases in order to draw a more complete picture of the stellar wind at different orbital phases \citep{Kretschmar_2021a}. 

We have confirmed that the photon index is anti-correlated with luminosity but cannot confirm the suggested correlation of the energy of the harmonic CRSF with flux. While the anti-correlation of the photon index with luminosity points towards subcritical accretion for Vela X-1, theoretical considerations for the critical luminosity are inconclusive and will require further investigation. The theoretical expectations for the evolution of the energy of the harmonic CRSF with luminosity shown in Fig.~\ref{fig:furst2014_fig7} are based (unavoidably) on rather simple, mostly fixed geometries. The intermediate case between two accretion regimes for Vela X-1 is the most difficult to treat, and it could be interesting in the future to explore more variability than height-scaling in theoretical studies, including full radiation hydro simulations as suggested by \citet{Vybornov_2017}.
Still, we observe variability in CRSF energy and in particular a drop following a flaring episode. Further observations of Vela X-1 and careful analysis of flaring episodes are necessary to test whether this observation was spurious or implies a change in the accretion geometry after a strong flare.

\begin{acknowledgements} We thank the anonymous referee whose input helped to strengthen the paper. This work has been partially funded by the  Bundesministerium f\"ur Wirtschaft und Energie under the Deutsches Zentrum f\"ur Luft- und Raumfahrt Grants 50~OR~1915.
SMN acknowledges funding under project RTI2018-096686-B-C21 funded by MCIN/AEI/10.13039/501100011033 and by 'ERDF A way of making Europe', and by the Unidad de Excelencia María de Maeztu, ref. MDM-2017-0765. ESL acknowledges support by DFG grant 1830Wi1860/11-1 and RFBR grant 18-502-12025. KP acknowledges support by NASA under award number 80GSFC17M0002.
This work has made use of (1) the Interactive Spectral Interpretation System (ISIS) maintained by Chandra X-ray Center group at MIT; (2) the \nustar Data Analysis Software (NuSTARDAS) jointly developed by the ASI Science
Data Center (ASDC, Italy) and the California Institute of Technology (Caltech, USA); (3) the of ISIS functions (\texttt{isisscripts})\footnote{\url{http://www.sternwarte.uni-erlangen.de/isis/}\label{footnote:isisscripts}} provided by ECAP/Remeis observatory and MIT; (4) NASA's
Astrophysics Data System Bibliographic Service (ADS). We thank John E. Davis for the development of the
\texttt{slxfig}\footnote{\url{http://www.jedsoft.org/fun/slxfig/}\label{footnote:jedsoft}}
module used to prepare most of the figures in this work. 
\end{acknowledgements}

\bibliographystyle{aa}
\bibliography{aa_abbrv,mnemonic,references}

\onecolumn

\begin{appendix}

\section{Additional tables}

\begin{table*}[!htpb]
\caption{\label{fig:app1}Best-fit parameters for the time-averaged tested models for observation I.}
\begin{center}
\begin{small}
\begin{tabular}{lllll}    
\hline\hline
Parameter & 
\texttt{FDcut} &
\texttt{FDcut} (w/o 10\,keV) &
\texttt{highecut} (w/o 10\,keV) \\
\hline

$C_{\rm{FPMA}}$  &  1  &  1  &  1  &  1\\
$C_{\rm{FPMB}}$  &  $1.0203\pm0.0021$  & $1.0203\pm0.0021$ & $1.0203\pm0.0021$ & $1.0204\pm0.0021$\\
$N_{\rm{H,1}} \ (10^{22} \,  \rm{cm^{-2}})$  &  $34.0\pm1.0$  &  $34.4\pm1.0$ & $36.4^{+1.0}_{-1.1}$ & $35.2^{+0.9}_{-1.0}$\\
$N_{\rm{H,2}} \ (10^{22} \, \rm{cm^{-2}})$  &  fixed to 0.371  &  fixed to 0.371  & fixed to 0.371 & fixed to 0.371\\
$\Gamma$  &  $1.09\pm0.05$  & $1.294^{+0.017}_{-0.033}$ & $1.373^{+0.015}_{-0.017}$ &$1.353\pm0.016$\\
$E_{\rm{cut}} \ (\rm{keV})$  &  $19.9^{+3.5}_{-2.0}$  &  $40^{+0}_{-5}$ & $25.6^{+0.5}_{-0.4}$ &$28.1^{+0.7}_{-1.0}$\\
$E_{\rm{fold}} \ (\rm{keV})$  &  $12.6^{+0.9}_{-0.8}$  &  $10.5^{+1.0}_{-0.7}$ & $15.0^{+0.7}_{-0.6}$ & $16.0^{+1.2}_{-1.0}$\\
$E_{\rm{CRSF,F}} \ (\rm{keV})$  &  $24.7^{+1.0}_{-0.9}$  & $27.38^{+0.26}_{-0.29}$ & $25.8^{+0.5}_{-0.4}$ & $27.6^{+0.5}_{-0.6}$\\
$\sigma_{\rm{CRSF,F}} \ (\rm{keV})$  & $0.5\times \sigma_{\rm{CRSF,H}}$  & $0.5\times \sigma_{\rm{CRSF,H}}$ & $0.5\times \sigma_{\rm{CRSF,H}}$ & $0.5\times \sigma_{\rm{CRSF,H}}$\\
$d_{\rm{CRSF,F}} \ (\rm{keV})$  &  $0.75^{+0.00}_{-0.19}$  & $2.82^{+0.15}_{-0.61}$ & $3.3\pm0.4$ & $4.9^{+0.4}_{-0.5}$\\
$E_{\rm{CRSF,H}} \ (\rm{keV})$  &  $53.8^{+1.1}_{-0.9}$  & $53.6\pm0.8$ & $54.4\pm0.5$ & $55.3\pm0.7$\\
$\sigma_{\rm{CRSF,H}} \ (\rm{keV})$  &  $7.9^{+1.3}_{-0.9}$  & $10.9^{+0.4}_{-0.8}$ & $8.7^{+2.0}_{-0.6}$ & $10.7^{+0.5}_{-0.7}$\\
$d_{\rm{CRSF,H}} \ (\rm{keV})$  &  $18^{+6}_{-4}$  & $38^{+4}_{-9}$ & $19.4^{+14.0}_{-2.9}$ &$33\pm6$\\
$E_{\rm{FeK\alpha}} \ (\rm{keV})$  &  $6.364\pm0.012$  & $6.365\pm0.012$ & $6.365\pm0.012$ &$6.367\pm0.012$ \\
$A_{\rm{FeK\alpha}} \ (\rm{ph \, s^{-1} \, cm^{-2}})$  &  $\left(1.35\pm0.11\right)\times10^{-3}$ & $\left(1.44\pm0.11\right)\times10^{-3}$ & $\left(1.39^{+0.13}_{-0.11}\right)\times10^{-3}$ & $\left(1.36\pm0.11\right)\times10^{-3}$\\
$\sigma_{\rm{FeK\alpha}} \ (\rm{keV})$  &  $0.070^{+0.029}_{-0.045}$  & $0.084^{+0.027}_{-0.033}$ & $0.07^{+0.04}_{-0.05}$ & $0.067^{+0.029}_{-0.053}$\\
$E_{10\,\rm{keV}} \ (\rm{keV})$  &  $9.5^{+0.6}_{-1.0}$  & -- & $8.00^{+0.27}_{-0.00}$ & -- \\
$A_{10\,\rm{keV}} \ (\rm{ph \, s^{-1} \, cm^{-2}})$  &  $\left(-4.8^{+2.2}_{-4.8}\right)\times10^{-3}$  & -- & $\left(4.8^{+2.1}_{-1.1}\right)\times10^{-4}$ & -- \\
$\sigma_{10\,\rm{keV}} \ (\rm{keV})$  &  $3.2^{+1.0}_{-0.7}$  & -- & $0.40^{+0.32}_{-0.11}$ & -- \\
$\mathcal{F}_{3-79 \ \rm{keV}} \ (\rm{keV \, s^{-1} \, cm^{-2}})$  &  $3.43^{+0.10}_{-0.05}$  & $3.91^{+0.05}_{-0.21}$  & $3.642^{+0.012}_{-0.053}$ & $3.92^{+0.10}_{-0.11}$\\
$\rm{CF}$  &  $0.883\pm0.005$  & $0.898\pm0.004$ & $0.8979^{+0.0031}_{-0.0030}$ &$0.901\pm0.004$\\
$\chi^2/\rm{dof}$  &  $613.37/456$  & $692.60/459$ & $617.16/456$ & $698.07/459$\\
\hline
\end{tabular}
\end{small}
\end{center}
\end{table*}

\longtab[2]{
\begin{landscape}
\begin{longtable}{llllll}
\caption{\label{fig:app2}Best-fit parameters for the time-averaged tested models for observation I.}\\
\hline\hline
Parameter & 
single \texttt{compTT} &
single \texttt{compTT} (w/o 10\,keV) &
double \texttt{compTT} &
\texttt{NPEX} (w/o 10\,keV) \\
\hline

$C_{\rm{FPMA}}$  &  1  &  1  &  1  &  1  &  1\\
$C_{\rm{FPMB}}$  &  $1.0203\pm0.0021$ & $1.0203\pm0.0021$ & $1.0203\pm0.0021$ & $1.0203\pm0.0021$ & $1.0203\pm0.0021$ \\
$N_{\rm{H,1}} \ (10^{22} \,  \rm{cm^{-2}})$  &  $37.0^{+1.3}_{-1.4}$ & $35.8\pm1.1$ & $35.5\pm1.2$ & $34.3\pm1.1$ & $34.3\pm1.0$ \\
$N_{\rm{H,2}} \ (10^{22} \, \rm{cm^{-2}})$  &  fixed to 0.371  &  fixed to 0.371  & fixed to 0.371 & fixed to 0.371 & fixed to 0.371\\
$\Gamma_1$  & -- & -- & -- & $-0.51\pm0.14$ & $-0.71\pm0.04$ \\
$\Gamma_{1,\mathrm{norm}}$ & -- & -- & -- & $0.164^{+0.039}_{-0.030}$ & $0.167\pm0.010$ \\
$\Gamma_2$ & -- & -- & -- & fixed to 2 & fixed to 2 \\
$\Gamma_{2,\mathrm{norm}}$ & -- & -- & -- & $\left(1.3^{+0.6}_{-0.5}\right)\times10^{-3}$ & $\left(1.07^{+0.11}_{-0.10}\right)\times10^{-4}$ \\
$E_{\rm{cut}} \, (\rm{keV})$ & -- & -- & -- & $6.33^{+0.38}_{-0.29}$ & $7.75^{+0.34}_{-0.30}$ \\
$\texttt{compTT}_{\mathrm{norm},1}$ & $0.0559^{+0.0032}_{-0.0028}$ & $0.0541^{+0.0024}_{-0.0023}$ & $0.052^{+0.009}_{-0.025}$ & -- & -- \\
$T_{0,1} \, (\rm{keV})$ & $0.94^{+0.05}_{-0.07}$ & $1.000^{+0.000}_{-0.008}$ & $1.000^{+0.000}_{-0.018}$ & -- & -- \\
$kT_1 \, (\rm{keV})$ & $8.3^{+0.5}_{-0.4}$ & $8.2\pm0.5$ & $6.9^{+0.4}_{-0.8}$ & -- & -- \\
$\tau_1$ & $10.43\pm0.22$ & $10.52^{+0.23}_{-0.22}$ & $11.9^{+1.2}_{-0.9}$ & -- & -- \\
$\texttt{compTT}_{\mathrm{norm},2}$ & -- & -- & $\left(6^{+25}_{-5}\right)\times10^{-3}$ & -- & -- \\
$T_{0,2} \, (\rm{keV})$ & -- & -- & fixed to $T_{0,1}$ & -- & -- \\
$kT_2 \, (\rm{keV})$ & -- & -- & $12.3^{+2.7}_{-4.2}$ & -- & -- \\
$\tau_2$ & -- & -- & $6.6^{+6.5}_{-1.6}$ & -- & -- \\
$E_{\rm{CRSF,F}} \ (\rm{keV})$  &  $26.29^{+0.31}_{-0.00}$ & $26.2\pm0.4$ & $24.9^{+0.8}_{-0.6}$ & $27.9\pm0.7$ & $26.62\pm0.29$ \\
$\sigma_{\rm{CRSF,F}} \ (\rm{keV})$  &  $0.5\times \sigma_{\rm{CRSF,H}}$  & $0.5\times \sigma_{\rm{CRSF,H}}$ & $0.5\times \sigma_{\rm{CRSF,H}}$ & $0.5\times \sigma_{\rm{CRSF,H}}$ & $0.5\times \sigma_{\rm{CRSF,H}}$\\
$d_{\rm{CRSF,F}} \ (\rm{keV})$  &   $2.7\pm0.7$ & $2.7\pm0.8$ & $0.85^{+0.44}_{-0.20}$ & $18.7^{+2.4}_{-3.0}$ & $3.6^{+0.8}_{-0.7}$ \\
$E_{\rm{CRSF,H}} \ (\rm{keV})$  &  $54.9\pm0.9$ & $54.9\pm1.0$ & $53.6^{+1.0}_{-0.8}$ & $51.7^{+1.2}_{-1.0}$ & $54.5\pm0.8$ \\
$\sigma_{\rm{CRSF,H}} \ (\rm{keV})$ & $11.0^{+0.9}_{-1.0}$ & $11.0^{+1.0}_{-1.1}$ & $7.6^{+1.3}_{-0.8}$ & $14.2^{+0.4}_{-0.0}$ & $11.5\pm0.8$ \\
$d_{\rm{CRSF,H}} \ (\rm{keV})$  &  $36^{+10}_{-9}$ & $35^{+11}_{-10}$ & $16.2^{+5.9}_{-3.0}$ & $83\pm10$ & $40^{+10}_{-8}$ \\
$E_{\rm{FeK\alpha}} \ (\rm{keV})$  & $6.368\pm0.012$ & $6.368\pm0.012$ & $6.368\pm0.012$ & $6.364\pm0.012$ & $6.365\pm0.012$ \\
$A_{\rm{FeK\alpha}} \ (\rm{ph \, s^{-1} \, cm^{-2}})$  & $\left(1.44^{+0.12}_{-0.11}\right)\times10^{-3}$ & $\left(1.37\pm0.11\right)\times10^{-3}$ & $\left(1.40\pm0.11\right)\times10^{-3}$ & $\left(1.36\pm0.11\right)\times10^{-3}$ & $\left(1.40\pm0.11\right)\times10^{-3}$ \\
$\sigma_{\rm{FeK\alpha}} \ (\rm{keV})$  & $0.084^{+0.027}_{-0.034}$ & $0.076^{+0.028}_{-0.037}$ & $0.082^{+0.027}_{-0.035}$ & $0.072^{+0.029}_{-0.042}$ & $0.078^{+0.028}_{-0.037}$ \\
$E_{10\,\rm{keV}} \ (\rm{keV})$  &  $8.39^{+0.26}_{-0.20}$ & -- & -- & $9.8^{+0.5}_{-0.8}$ & -- \\
$A_{10\,\rm{keV}} \ (\rm{ph \, s^{-1} \, cm^{-2}})$  &  $\left(5.3^{+3.1}_{-2.4}\right)\times10^{-4}$ & -- & -- & $-0.37\pm0.11$ & -- \\
$\sigma_{10\,\rm{keV}} \ (\rm{keV})$  & $0.71^{+0.23}_{-0.20}$ & -- & -- & $6.6^{+0.5}_{-0.4}$ & -- \\
$\rm{CF}$  & $0.842^{+0.014}_{-0.012}$ & $0.831^{+0.005}_{-0.004}$ & $0.830\pm0.005$ & $0.878\pm0.010$ & $0.882\pm0.005$ \\
$\chi^2/\rm{dof}$  & $604.45/456$ & $630.30/459$ & $622.53/456$ & $604.15/456$ & $644.55/459$ \\
\hline
\end{longtable}
\end{landscape}
}

\end{appendix}

\end{document}